\documentclass[letterpaper]{article} % DO NOT CHANGE THIS
\usepackage{aaai24}  % DO NOT CHANGE THIS
\usepackage{times}  % DO NOT CHANGE THIS
\usepackage{helvet}  % DO NOT CHANGE THIS
\usepackage{courier}  % DO NOT CHANGE THIS
\usepackage[hyphens]{url}  % DO NOT CHANGE THIS
\usepackage{graphicx} % DO NOT CHANGE THIS
\usepackage{natbib}  % DO NOT CHANGE THIS AND DO NOT ADD ANY OPTIONS TO IT
\usepackage{caption} % DO NOT CHANGE THIS AND DO NOT ADD ANY OPTIONS TO IT

\usepackage{multirow}
\usepackage{appendix}
\usepackage{amsfonts}
\usepackage{epstopdf}
\usepackage{amsmath}
\usepackage{booktabs}

\usepackage{algorithm}
\usepackage{algorithmic}
\usepackage[table,xcdraw]{xcolor}

\usepackage{newfloat}
\usepackage{listings}
\DeclareCaptionStyle{ruled}{labelfont=normalfont,labelsep=colon,strut=off} % DO NOT CHANGE THIS
\floatstyle{ruled}
\newfloat{listing}{tb}{lst}{}
\floatname{listing}{Listing}

\newcommand{\RN}[1]{%
	\textup{\lowercase\expandafter{\it \romannumeral#1}}%
}

\nocopyright 
\title{InstructME: An Instruction Guided Music Edit And Remix Framework with Latent Diffusion Models}
\author{
    Bing Han\equalcontrib \textsuperscript{\rm 1} \thanks{Interns at ByteDance.}, Junyu Dai\equalcontrib \textsuperscript{\rm 2}, Weituo Hao\textsuperscript{\rm 2}, Xinyan He\textsuperscript{\rm 2}, Dong Guo\textsuperscript{\rm 2},  \\ Jitong Chen\textsuperscript{\rm 2}, Yuxuan Wang\textsuperscript{\rm 2}, Yanmin Qian\textsuperscript{\rm 1}, Xuchen Song\textsuperscript{\rm 2}
}
\affiliations{
    \textsuperscript{\rm 1}Department of Computer Science and Engineering, Shanghai Jiao Tong University\\
    \textsuperscript{\rm 2}ByteDance \\
    xuchen.song@bytedance.com
}

\usepackage{bibentry}
\begin{document}

\maketitle

\begin{abstract}
% 应用前景，需要提一句remix和instr editing，并展示其承接关系
%音乐编辑通常指对音乐进行乐器轨道层面的操作或者对音乐整体进行remix,应用广泛但有较高门槛，最近，由于diffusion模型具有强大的生成能力，让实现基于diffusion的低门槛音乐编辑工具成为可能

% comment: 前两句这么改怎么样： Manipulating music, such as music editing that modifies individual instrument tracks and music remixing that offers a fresh perspective through complex operations, typically demands a deep understanding of music domain knowledge. By leveraging the diffusion model, we proposed methods that uses language to introduce and control musical properties during audio manipulation. Our method places emphasis on the harmony and consistency of audio, which are especially crucial in music manipulation.

% Music editing usually refers to manipulating music on their
% instrument tracks, while remixing can give a new interpre-
% tation of the music through more complex operations. They
% have a wide range of application scenarios but require vast
% professional knowledge.
Music editing primarily entails the modification of instrument tracks or remixing in the whole, which offers a novel reinterpretation of the original piece through a series of operations. These music processing methods hold immense potential across various applications but demand substantial expertise.
% Intelligent music editing tools have a wide range of application scenarios to engage non-professional users to achieve creative manipulation of music. 
% Recently,  due to the unique nature of music data, previous editing work are difficult to directly apply to music. 
% 现有的方法不能用
% However, benefiting from the powerful generative capabilities of the diffusion model, it is possible now to implement a low-threshold music editing and remixing tool based on diffusion model. There have been some edit-related methods that work well on image and audio, but these methods will disrupt the harmony and consistency of music when applied to music directly, owing to the unique nature of music data.
Prior methodologies, although effective for image and audio modifications, falter when directly applied to music. This is attributed to music's distinctive data nature, where such methods can inadvertently compromise the intrinsic harmony and coherence of music.
% 提出了InstructME
In this paper, we develop InstructME, an \textbf{Instruct}ion guided \textbf{M}usic \textbf{E}diting and remixing framework based on latent diffusion models. 
% MSA
Our framework fortifies the U-Net with multi-scale aggregation in order to maintain consistency before and after editing.
% chord condition
In addition, we introduce chord progression matrix as condition information and incorporate it in the semantic space to improve melodic harmony while editing.
% CC
% To endow the InstructME with the ability to handle long music, chunk transformer is employed to model long-term temporal dependencies of music sequence in a chunk-wise manner. 
For accommodating extended musical pieces, InstructME employs a chunk transformer, enabling it to discern long-term temporal dependencies within music sequences.
% Summary
% Applying InstructME to specific tasks, such as instrument-editing, remixing and multi-round editing, and conducting subjective and objective experiments separately, the results show that InstructME outperforms the previous system by a large margin in terms of music quality, text relevance and harmony. 
We tested InstructME in instrument-editing, remixing, and multi-round editing. Both subjective and objective evaluations indicate that our proposed method significantly surpasses preceding systems in music quality, text relevance and harmony.
Demo samples are available at \url{https://musicedit.github.io/}
\end{abstract}

% \WT{help us to think through:}
% \begin{itemize}
%     \item What figure: 
% This figure shows only two things: input and output
% This helps the readers understand what your work is about.
%     \item Why figure:
% This figure provides with the “motivation” for your work. The best WHY figure illustrates the existing work’s core issue/problem with a concrete example.
%     \item How figure:
% This figure describes how your method works.
% Tips: Link all sections; Use consistent notations; self-contained caption; Visualize the variables; small units: “input - some processing - output” 
% \end{itemize}

\section{Introduction}

Music editing involves performing basic manipulations on musical compositions, including such atomic operations as the inclusion or exclusion of instrumental tracks and the adjustment of pitches in specific segments. On top of these atomic operations, remixing can be understood as an advanced version of music editing that mixes various atomic operations with style and genre considered~\cite{fagerjord2010after}.
Both atomic operations and remix can be handled by using a text-based generative model. In music editing, the text would be natural language-based editing instructions, such as \textit{"adding a guitar track"}, \textit{"replacing a piano track with a violin"}, etc. Models from the text-generated image domain seem to be adaptable to the music editing scenario. However, unlike image generation, models need to pay attention to the music harmony in addition to understanding the text and generating it. For example, when introducing a guitar track, care is taken to harmonize its rhythm, chord progression, and melodic motifs with the original audio framework, thus ensuring that overall consistency and coherence are maintained. Therefore, for successful music editing, the model should be able to:$(\RN{1})$ understand editing instructions and generate music stems;$(\RN{2})$ ensure the compatibility of the part being processed with the original music source.

\begin{figure}[h]
\centering
\includegraphics[width=1.0\columnwidth]{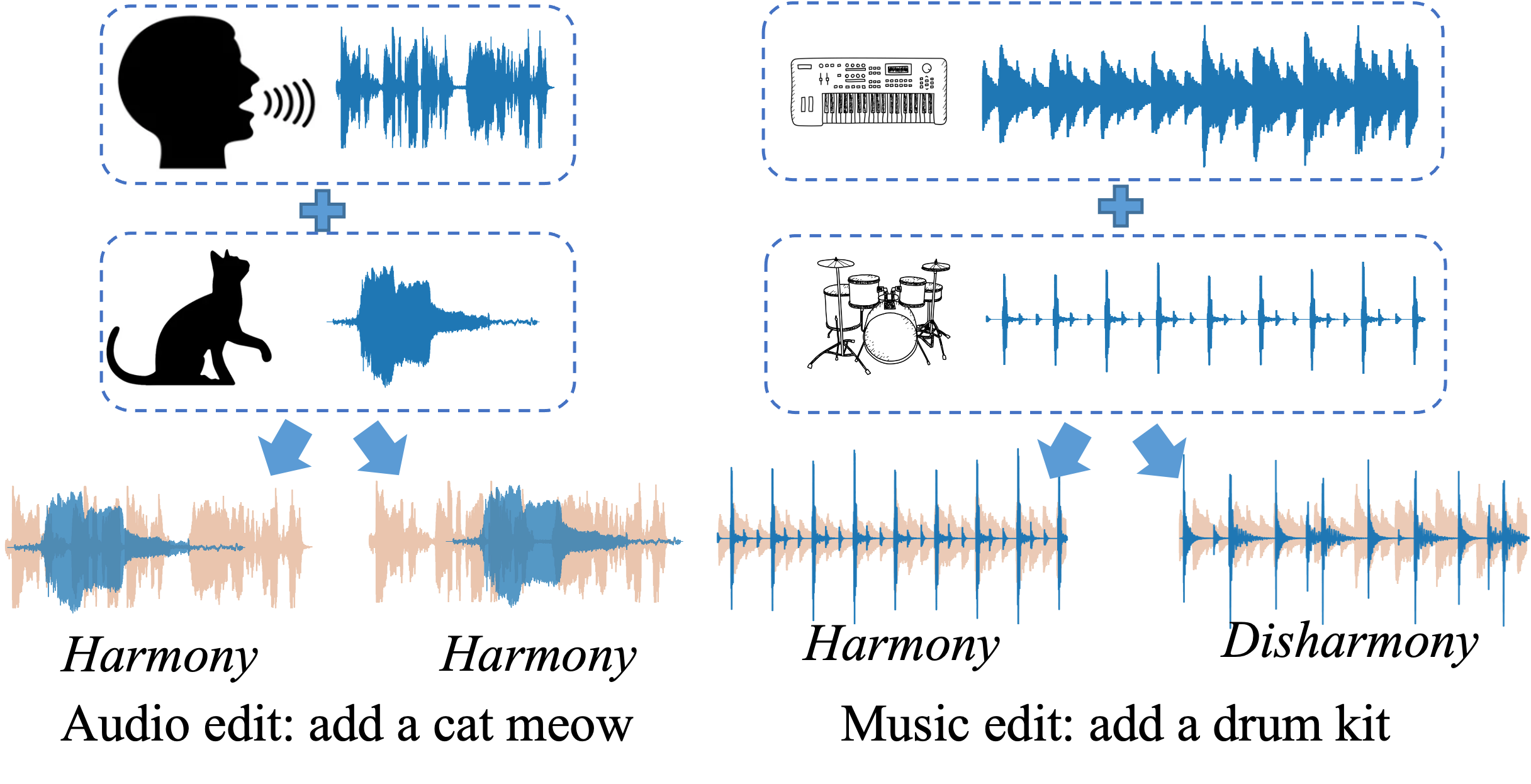}
\caption{Left is audio edit: Each audio component is independent that does not necessitate the consideration of interdependence. Right is music edit: Harmony in pitch, intensity, rhythm, and timbre must be taken into account.}
%(a): Image editing explicitly modifies the area to be edited by employing masking techniques. (b) Each audio component in audio editing is independent that does not necessitate the consideration of interdependence. (c) Harmony in rhythm and timbre must be taken into account when editing music.
% The components of music overlap in both the time and frequency domains, which leads to the utilization of uncomplicated masks to edit music being impeded.
\label{fig:why_figure}
\end{figure}

Lately, a multitude of endeavours pertaining to text-based image or audio manipulation\cite{hertz2022prompt, lugmayr2022repaint, meng2021sdedit, wang2023audit} have attracted considerable attention due to their noteworthy performance within their respective domains. However, the distinct data properties and generative prerequisites inherent to the domain of music preclude the direct applicability of these methods to the sphere of music editing. In image editing, it is feasible to maintain consistency over the residual regions by employing masking techniques, thereby confining attention solely to the objects to be generated. However, this underlying principle proves inapplicable to the domain of musical data, as shown in Figure~\ref{fig:why_figure} the interwoven nature of individual tracks across both temporal and frequency domains prevents the straightforward implementation of such an approach. Perhaps the most similar method to ours is ~\cite{wang2023audit} for audio editing. However, the method is mainly applied to the editing of sound effects. Unlike music tracks, the individual sound effects are independent of each other, so there is no need to consider whether the sound effects are in harmony with each other or not.

%将任务统一，并针对目前框架在音乐编辑上的缺点，我们提出instrME，他在audit的基础上针对音乐相关任务做了一些改进
%为了解决现有框架应用于音乐编辑任务带来的不和谐、不连续问题，我们引入instrme，一个基于ldm的music edit framework. 需要说明的是，本文将基础操作限定在对乐器的增删查改上，重点考虑操作前后音乐的和谐和连续性，而将reverb、pitch shift等常规音乐调整排除在外，这是由于使用传统dsp的方式实现这些操作就已经足够方便且performance well。同时，我们将对音乐元素的基础编辑和remix进行统一，将remix作为基础编辑的组合，同时二者统一为一种更普遍编辑任务的实例
\begin{figure}[h]
\centering
\includegraphics[width=1.0\columnwidth]{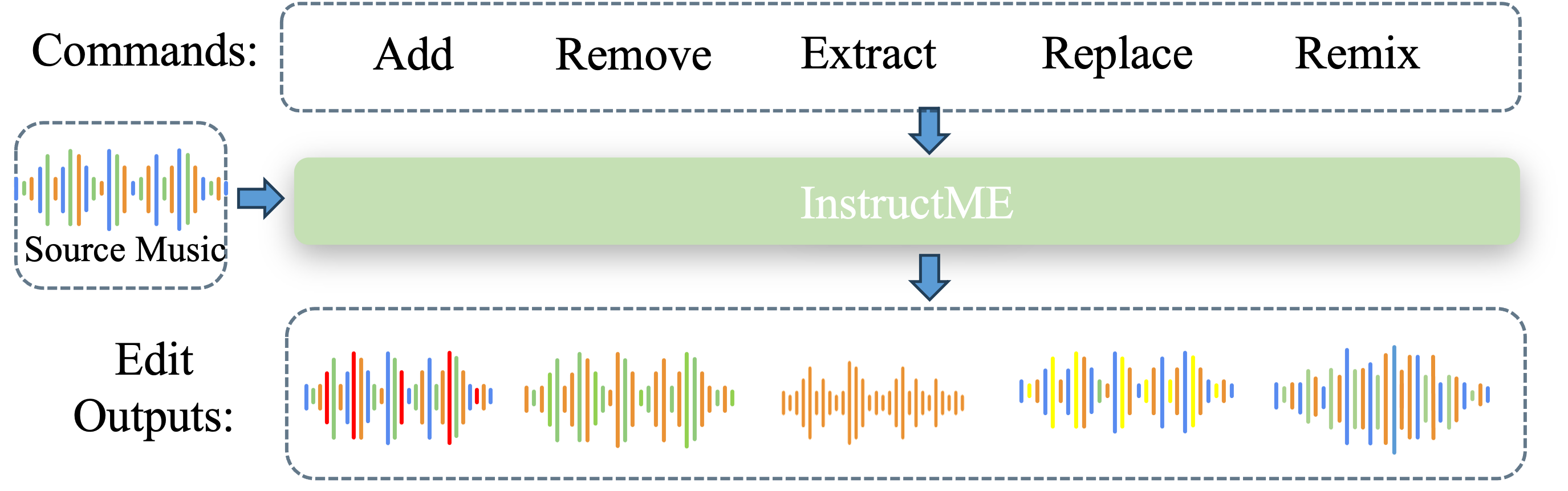} % Reduce the figure size so that it is slightly narrower than the column. Don't use precise values for figure width. This setup will avoid overfull boxes.
\caption{A brief illustration of InstructME. Given source music and command text, InstructME generates a piece of music that is harmonious and complies with the command requirement.}
% It provides an easy approach to manipulating or arranging the music tracks.
\label{fig:what_figure}
    \vspace{-0.2cm}
\end{figure}
In order to bridge the gap between text-based generative models and music editing tasks, we propose InstructME, an instruction-guided music editing framework based on latent diffusion models. For simplicity, we limit music editing operations to adding, removing, extracting, replacing, and remixing. %a combination of these, remixing. 
% Conventional music manipulating such as adding reverb and pitch shifting are excluded from consideration, as these operations can be conveniently and efficiently implemented using traditional DSP (Digital Signal Processing) methods. 
%如图所示，instrme是这样一个framework，给定一个原始音乐以及操作文本，不论是基础操作还是针对乐器或者风格的remixing，instrme将根据操作文本对原始音乐进行编辑，并使得编辑结果具有音乐和谐性与一致性。 
%, whether it is basic operations or remixing,
% As shown in Figure \ref{fig:what_figure}, InstructME is the framework that given source music and command text data pair, can perform editing on the source music based on the command text and can ensure that the edited result is harmony and consistency. InstructME takes AUDIT~\cite{wang2023audit} as a starting point, but different from AUDIT and inspired by ~\cite{karras2020analyzing}, 
As shown in Figure \ref{fig:what_figure}, InstructME takes text instructions and source music as input, and outputs the target music accordingly. To maintain the consistency of the music before and after editing, we utilize the multi-scale aggregation strategy and incorporate the chord progression matrix into the semantic space~\cite{kwon2022diffusion,jeong2023training} during the source music encoding process to ensure harmony. During training, we employ the chunk transformer to model long-term temporal dependencies of music data in a segmented chunk-wise manner and train the model on collected 417 hours of music data.
%({command, source music, target music})
For testing, we evaluate the model in terms of three aspects: music quality, text relevance and harmony. Experimental results of public and private datasets demonstrate that InstructME outperforms the previous system.
% although it is difficult to quantitatively analyze the performance of music editing, after referring to \cite{lv2023getmusic, ren2020popmag, kilgour2019frechet, yang2020evaluation}, 
%而在评估上，尽管音乐编辑的表现很难定量评估，但参考了这些工作后，我们尝试从音乐质量、文本相关性和音乐和谐度三个方面进行讨论，分别给与定量指标并设计实验测试，从实验结果看，InstructME outperforms the previous system
%此外，与audit一致，我们构建了总共几百小时、千万个三元组数据pair，并在前后端接入autoencoder的encoder和decoder，以保证diffusion模型在latent空间进行训练
% % In this paper, we formulate Music Remixing as an instance of a more general \textbf{Music Editing} task, whereby we manipulate individual elements to transform existing music into new versions with targeted properties adhering and not losing connection with the vocal or some other recognizable tracks of the original recordings.

Our key contributions can be summarized as:
\begin{itemize}
    \item To the best of our knowledge, we propose the first instruction guided music editing framework applicable for both atomic and advanced operations. %This is the first work that focuses on music editing tasks especially remixing tasks as we know. 
    \item We point out the special problem of consistency and harmony in music editing domain and develop multi-scale aggregation and chord condition via chunk transformer to solve it.
    \item We propose quantitative evaluation metrics for music editing tasks in terms of music quality, text relevance and harmony.
    \item Our proposed method InstructME surpasses previous systems through thorough subjective and objective tests.
\end{itemize}

\section{Related Work}

\subsection{Text guided Generation}
Generating a new version of accompaniment for a track directly with targeted properties (e.g. genre, mood, instruments) adhering is a viable approach to accomplish the objectives of editing or remixing. Recent studies ~\cite{huang2023make, liu2023audioldm, agostinelli2023musiclm, schneider2023mo, huang2023noise2music, lam2023efficient, copet2023simple} have already succeeded in generating plausible music that reflects key music properties(e.g. genre, mood, etc) that are depicted in a given text. However, there is no guarantee for them to generate tracks that are harmonious with a given track while keeping the given one or specified part of it unchanged. Another work ~\cite{donahue2023singsong} proposed a generative model, which trained over instrumentals given vocals, generating coherent instrumental music to accompany input vocals. But it has no way for users to control the generation process, not to mention interactive editing, which is important for an intelligent editing tool as it applies feedback from users to make a more preferable output as in ~\cite{midjourney2023}.
%Interactive editing ability is important for intelligent content creation tools as it applies feedback from users to make a more preferable output as in ~\cite{midjourney2023}. 

\subsection{Audio Editing and Music Remixing}

~\cite{huang2023make, liu2023audioldm} propose zero-shot audio editing by utilizing pre-trained text-to-audio latent diffusion models, which seem flexible but not accurate enough for the editing process.
% Though at a glance, using free-form text or unlimited prompts as guidance in the interaction process seems flexible, in the content editing scenario, using a set of instructions as guidance instead can provide a more precise meaning of operations, leading to a more accurate and efficient interactive editing process.
Moreover, there is no guarantee for those audio generation models that are trained with general purposes to achieve a good editing effect in the editing specialized usage scenario. Due to this, AUDIT ~\cite{wang2023audit} proposed a general audio editing model based on a latent diffusion and denoising process guided by instructions. Certainly, as previously stated in the introduction section, this framework necessitates certain enhancements to effectively cater to music-related tasks.

For remixing, although the text-guided generative systems mentioned above can also perform generation conditions on a given recording ~\cite{liu2023audioldm, lam2023efficient}, or more specifically, melodies ~\cite{agostinelli2023musiclm, copet2023simple}, the generated music can only preserve the tune of the conditional melodies, the original tracks, such as vocal, will not directly feature in the output music. This kind of conditional-generated music is traditionally known as music covers, not remixes. Likewise, past studies ~\cite{yang2022don, yang2020remixing, wierstorf2017perceptual} have attempted to apply neural networks to the task of music remixing. These methods, which are often incorporated with source separation models, primarily viewed music remixing as a task of adjusting the gain of individual instrument sources of an audio mixture. But music remixing is not just limited to manipulating the gain of different sources of the recording itself, it can also involve incorporating other materials to create something new ~\cite{waysdorf2021remix}. 

\begin{figure*}[h]
\centering
\includegraphics[width=2.0\columnwidth]{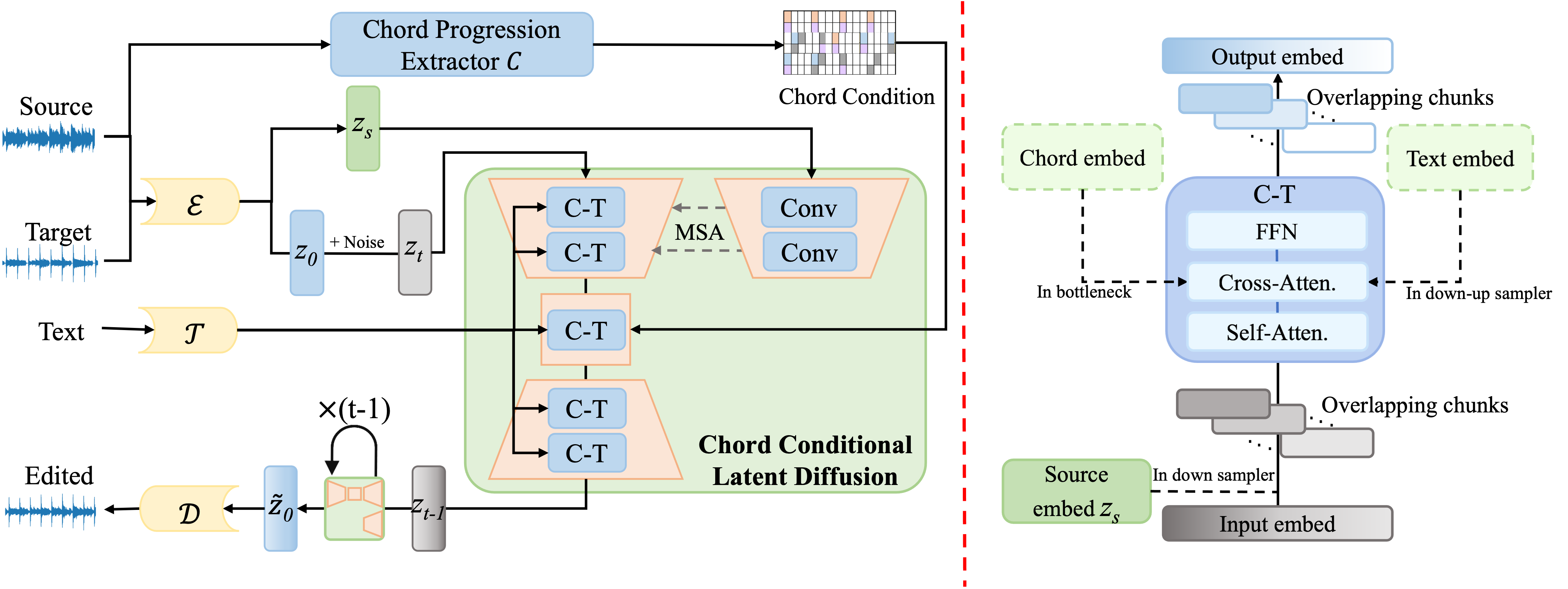} % Reduce the figure size so that it is slightly narrower than the column. Don't use precise values for figure width.This setup will avoid overfull boxes.
\caption{Left: Overview of InstructME diffusion process for music editing. Audio signal is processed by VAE (encoder$\mathcal{E}$ and decoder$\mathcal{D}$ ), meanwhile extractor($\mathcal{C}$) extracts the chord matrix of source music and together with text embedding extracted by $\mathcal{T}$ as condition information, latent embedding $z_{s}$ and $z_{t}$ are fused by multi-scale aggregation and converted by chunk transformer to produce the final edited music. Right: Architecture of chunk transformer(C-T) blocks which in various positions of U-net will selectively incorporate chord or text embedding, and $z_{s}$ will only input when chunk transformer is in down sampler.}
\label{fig:how_figure}
    \vspace{-0.2cm}
\end{figure*}

\section{Methodology}
In this section, we will provide an overview of the InstructME architecture and the process of instruction-based music editing, as illustrated in Figure~\ref{fig:how_figure}. Additionally, we will explain strategies aimed at improving editing consistency and harmony, as well as approaches to achieving more sophisticated music editing operations.

% We provide an overview of InstructME model architecture in Figure~\ref{fig:how_figure}. In this section, we will go through text-based music generation steps, and approaches that improve generation consistency and achieve advanced music editing operation respectively.
% , which consists of several parts: chord-conditional latent diffusion model, multi-scale aggregation and chunk transformer. 

\subsection{Instruction To Music Editing}
InstructME accepts music audio $\mathbf{x}_{s}$ and editing instructions $y$ as input, and produces new audio $\mathbf{x}$  that adheres to the given instructions. We utilize text and audio encoders to transform the data into a latent representation. For each text instruction $y$, a pretrained T5~\cite{raffel2020exploring} converts it into sequence of embeddings $\mathcal{T}(y)\in \mathbb{R}^{L \times D}$,  similar to~\cite{wang2023audit}. For each audio segment $\mathbf{x}_{s}\in \mathbb{R}^{T\times 1}$, a variational auto-encoder (VAE) transforms the waveform into a 2D latent embedding $\mathbf{z}_{s}\in \mathbb{R}^{\frac{T}{r} \times C }$. Using text and audio embeddings as conditions, a diffusion process ~\cite{song2020score, ho2020denoising} produces embeddings of new audio samples, which the VAE decoder then converts back to audio waveforms. 

%Unlike image and text modalities, audio often has extremely low information density and ultra long sequences. %For example, in 24khz sample rate, music with 10 seconds contains 240,000 samples, which is approximately 234 images in 32x32.
%To lower the computational demands of training diffusion models towards efficient music editing, we employ an autoencoder to learn a low-dimensional latent space which is perceptually equivalent to the waveform space, but offers significantly reduced computational complexity, like latent diffusion~\cite{rombach2022high}.

The VAE used by InstructME consists of an encoder $\mathcal{E}$, a decoder $\mathcal{D}$ and a discriminator with stacked convolutional blocks. The decoder reconstructs the waveform $\hat{\mathbf{x}}$ from the latent space $\mathbf{z}$ and there is no vocoder like~\cite{wang2023audit,liu2023audioldm}. The discriminator was used to enhance the sound quality of generated audio through adversarial training. We provide the model and training details in the Appendix.

\subsubsection{Diffusion Model}
% In InstructME, we employ diffusion models~\cite{song2020score, ho2020denoising} as the probabilistic generative model to achieve the ability to generate the edited music $x_{tgt}$ given source music $x_{s}$ and text description $y$. 
% And this problem can be formulated as:
% \begin{equation}
% \label{equ:model}
%     x_{tgt} = \mathcal{M}(x_{s}, y)
%     % \nonumber
% \end{equation}
% where $\mathcal{M}$ denote the diffusion based music edit model.

Diffusion model contains two processes. The forward process is a standard Gaussian noise injection process. %In the forward process, it transforms the data distribution into standard Gaussian distribution by adding Gaussian noise step by step with a predefined noise schedule, $0<\beta_1 < \dots < \beta_t < \dots < \beta_T$. And a
At time step $t$, 
% the transition probability is:
% \begin{equation}
%     \label{equ:transition_prob}
%     q(z_{t} | z_{  t-1}) = \mathcal{N}(z_{ t} ; \sqrt{1-\beta_t}z_{ t-1}, \beta_t \mathbf{I}) 
%     % \nonumber
% \end{equation}
% where $z=\mathcal{E}(x_{tgt})$ is the latent representation of target music, encoded by autoencoder $\mathcal{E}$. Then, according to the Chain rule of probability,
\begin{align}
    % \nonumber
    q(z_{t} | z_{0}) %&= \prod_{t=1}^T q(z_{t} | z_{t-1}) \\ 
    & = \mathcal{N}(z_{t} ; \sqrt{\bar{\alpha}_t}z_{0}, (1 -\bar{\alpha}_t) \epsilon) 
    % \nonumber
\end{align}
where $\alpha_t = 1-\beta_t$ and $\bar{\alpha}_t=\prod^T_{t=1}{\alpha_t}$ are scheduling hyperparameters. %Noted that $\epsilon$ here refers to the standard Gaussian noise which is injected on clean latent $z_{ 0}$. 

In the reverse process, we employ a time-conditional U-Net $\epsilon_{\theta}$~\cite{ronneberger2015u,rombach2022high} as the denoise model backbone. At time step $t$, conditioning on embeddings of text $\mathcal{T}(y)$ and source music $z_{s}$, this denoise model attempts to restore the original latent $z_{0}$ of target music from noisy $z_{t}$. For model optimization, we use reweighted bound~\cite{ho2020denoising,rombach2022high} as objective function:
\begin{equation}
    \label{equ:diffusion_loss}
    \mathcal{L}_{DM}= \mathbb{E}_{\epsilon,t,z_{ 0}} \left \| \epsilon-\epsilon_{\theta}(t, \mathcal{T}(y), z_{s},z_{ t}) \right \| _2^2
\end{equation}
with $t$ uniformly sampled from $[1,T]$ during the training. In the end, we pass $z_0$ through the decoder $\mathcal{D}$ to obtain the waveform music.

The U-Net layers utilize transformers with self and cross attention as building blocks. In the down sampler layers, source audio embeddings and generated embeddings merge into the input of the self-attention layer. Cross attention is employed for text conditions in each down-up sampler layer, and for chord conditions in the bottleneck layer.

\subsubsection{Efficient Diffusion}
The self-attention of lengthy music sequences is computationally expensive. To alleviate this problem, we employ the chunk transformer to model long-term temporal dependencies in a chunk-wise manner. Outlined in Figure 3 (Right), the process involves three steps: segmentation of $T$-frame embeddings into $K$-frame chunks with 50\% overlap, individual chunk processing through a transformer layer, and fusion to merge overlapping output chunks into $T$ frames.

% %For a 2D music sequence of $T$
% %$\mathbf{H}\in \mathbb{R}^{F\times T}$,
% frames the segment module can chunk the sequence into several non-overlapping segments with length $K$, denoted as $\mathbf{C_0}=\{c^0_0, \dots, c^S_0\} \in \mathbb{R}^{S\times F\times K}$, where $S=\left \lceil \frac{T}{K} \right \rceil$ is the number of segments and $c_0^k=\mathbf{H}[:,k*K:(k+1)*K)]$. 
% % padding
% In order to model the relationship among segments, similar to~\cite{liu2021swin}, we shift the sliding window and split the input sequence into $\mathbf{C_1}=\{c^0_1, \dots, c^S_1\}$, where $c_1^k=\mathbf{H}[:,\left \lfloor \frac{2k-1}{2}\right \rfloor *K:\left \lfloor \frac{2k+1}{2}\right \rfloor*K]$. Then, $\mathbf{C_0}$ and $\mathbf{C_1}$ are concatenated as $\mathbf{C}\in \mathbb{R}^{2*S\times F \times K}$ and fed to attention blocks to model temporal dependency. After that, segments will be merged and restored to the original sequences. By fusing two sequences, each token can see adjacent $\frac{3K}{2}$ tokens. 

At each layer of the chunk transformer, a token from a $K$-frame chunk can observe $\frac{3K}{2}$ neighboring frames. By stacking multiple layers of  chunk transformer, the U-Net acquires an expansive receptive field, enabling effective modeling of long-term dependencies. Compared with oracle transformer's complexity $\mathcal{O}(T^2)$, chunk transformer has lower computational cost $\mathcal{O}(2 * \left \lceil \frac{T}{K} \right \rceil * K^2) = \mathcal{O}(TK)$. 
In addition to faster inference and lower memory consumption, the chunk-wise modeling approach decreases the model's reliance on sequence length, learning invariant representations. This minimizes performance degradation caused by duration differences in training and sampling.

% \begin{equation}
%     \label{equ:complexity}
%     % \nonumber
%     \mathcal{O}(2 * S * K^2) = \mathcal{O}(2 * \left \lceil \frac{T}{K} \right \rceil * K^2) = \mathcal{O}(TK)
% \end{equation}
% which can improve speed and reduce memory usage during training and sampling. 

\subsection{Improving Consistency and Harmony}
To make the diffusion model more suitable for music editing tasks, we propose an enhanced U-Net with several modifications including multi-scale aggregation and chord condition.
\subsubsection{Multi-Scale Aggregation}
Contrary to the music generation tasks~\cite{huang2023noise2music}, music editing tasks require the preservation of certain content and properties from the original music. In order to maintain coherence between the original and edited music, AUDIT~\cite{wang2023audit} directly concatenates the source music channel $z_t$ with the target music channel $z_s$ at the U-Net's input. It leans heavily on the invariance of some low-level and local music features, which might pose challenges or limitations when applied to more complex music manipulation tasks. To more effectively capture the high-level characteristics of the source music, we introduce a multi-scale aggregation (MSA) strategy
as depicted in Figure~\ref{fig:how_figure} (Left).
% Different from traditional generative tasks~\cite{huang2023noise2music}, editing tasks need to ensure that most of the content remains unchanged during the editing process. Different from AUDIT~\cite{wang2023audit} which takes source music as a condition by directly concatenating $z_t$ and $z_s$ at the channel level, to better model the source music, we propose the multi-scale aggregation strategy to incorporate a progressive path in U-Net, as shown in Figure~\ref{fig:how_figure}.
% It feeds the source music embeddings $z_s$ to a multi-layer convolution encoder, and provides the feature map with different resolutions to the corresponding U-Net layers, which is demonstrated to be effective in high-resolution image generation~\cite{karras2020analyzing}. 
The source music embeddings $z_s$ are input to a multi-layer convolution encoder, yielding feature maps with varying resolutions for the corresponding U-Net layers. This strategy has been proven effective in high-resolution image generation~\cite{karras2020analyzing}.
%It should be noted that this progressive path consists of several convolutional layers to implement downsampling operation and they don't use shared weights with U-net blocks.

\subsubsection{Chord-Conditional}
% Chord progression is an important aspect of measuring musical harmony. To improve the melodic coherence between source and edited music
The Chord progression is a key element in defining a piece's musical harmony.
% We utilize it to enhance the melodic consistency between the source and edited music, as shown in Figure~\ref{fig:how_figure},
We adopt a chord progression recognition model $\mathcal{C}$~\cite{cheuk2022jointist} to extract the chord probability embedding $p$ of the source music and then emphasize it explicitly during the denoise process. ~\cite{kwon2022diffusion,jeong2023training} discover the semantic latent space in the bottleneck of diffusion has nice properties to accommodate semantic image manipulation. Inspired by them, we incorporate the chord progression representation $p\in \mathbb{R}^{d_p \times T_p}$ in the bottleneck feature map $h\in \mathbb{R}^{d_h \times T_h}$ of U-Net with cross-attention mechanism~\cite{vaswani2017attention}.
% :$\mathrm{Attention}(Q,K,V)=\mathrm{Softmax}(\frac{Q^\mathrm{T}K}{\sqrt{d}})\cdot V^\mathrm{T}$,
% \begin{equation}
%     % \nonumber
%     \mathrm{Attention}(Q,K,V)=\mathrm{Softmax}(\frac{Q^\mathrm{T}K}{\sqrt{d}})\cdot V^\mathrm{T}
% \end{equation}
% with $Q=W_Q \cdot h$, $K=W_K \cdot p$ and $V=W_V \cdot p$, where $W_Q \in \mathbb{R}^{d\times d_p}$, $W_K \in \mathbb{R}^{d\times d_h}$, $W_V \in \mathbb{R}^{d\times d_h}$ are trainable linear transformation matrices. 
With chord progression condition extracted by $\mathcal{C}$, in the bottleneck layer of U-Net, the objective function in Equation~\ref{equ:diffusion_loss} can be rewritten as:
\begin{equation}
    \label{equ:diffusion_loss2}
    \mathcal{L}_{CDM}= \mathbb{E}_{\epsilon,t,z_{ 0}} \left \| \epsilon-\epsilon_{\theta}(t, p_{s}, z_{s},z_{ t}) \right \| _2^2
    % \nonumber
\end{equation}
where $p_{s}$ denotes chord progression matrix of source music $x_s$, encoded by $\mathcal{C}$.

% \cite{lam2023efficient}

% chunk transformer~\ref{fig:chunk_transformer}.

%\subsection{Guided Remix}
\subsection{Towards Advanced Music Editing - Remix}
% 对于guided diffusion generation，一般有两种方法。classifier-free 不需要重新训练模型，但依赖于大规模的多样性数据来作为text condition；classifier
For diffusion models, there exist two primary strategies for achieving controllable generation. One of these is classifier guidance (CG)~\cite{dhariwal2021diffusion,liu2023more}, which utilizes a classifier during the sampling process and mixes its input gradient of the log probability with the score estimate of diffusion model. It is flexible and controllable, but tends to suffer a performance degradation~\cite{ho2022classifier}. Another approach, named classifier-free guidance (CFG)~\cite{ho2022classifier,nichol2021glide,ramesh2022hierarchical,saharia2022photorealistic}, achieves the same effect through training a conditional diffusion model directly without a guidance classifier. This method performs better but requires a large amount of data with diverse text descriptions, which is difficult for our InstructME trained with source-target paired data. In this work, to attain a tradeoff between quality and controllability, we adopt both classifier and classifier-free guidance to achieve the controllable editing of Remix operations.

% For instrument, genre and other attributes are easy to be collected and labelled, 
We specify instrument and genre tags with CFG by incorporating these tags into text commands to train the conditional diffusion models. During the training, we discard our text condition $y$ randomly with a certain probability $p_{\mathrm{CFG}}$ following~\cite{liu2023audioldm,wang2023audit}. Then, in the sampling, we can estimate the noise $\hat{\epsilon}_{\theta}(t, \mathcal{T}(y), p_{s}, z_{s},z_{t})$ with a linear combination of the conditional and unconditional score estimates:
% \begin{equation}
\begin{align}
\nonumber
    \hat{\epsilon}_{\theta}(t, \mathcal{T}(y), p_{s}, z_{s},z_{ t}) & = (1 - w) \epsilon_{\theta}(t, p_{s}, z_{s},z_{ t})  \\ 
    & +  w\epsilon_{\theta}(t, \mathcal{T}(y), p_{s}, z_{s},z_{ t})
\end{align}
% \end{equation}
where $w$ can determine the strength of guidance.

To achieve finer-grained semantic control with weakly-associated, free-form text annotations, we apply classifier guidance during sampling with a pre-trained MuLan~\cite{huang2022mulan}, which can project the music audio and its corresponding text description into the same embedding space. The guidance function we use is:
\begin{equation}
    F(x_t, y) = \left \| E_{L}(y) - E_{M}(x_t) \right \| _2^2
\end{equation}
where $E_{L}(\cdot)$ and $E_{M}(\cdot)$ denote the language and music encoders respectively. Then, by adding the gradient on estimated $x_t$, we can guide the generation 
\begin{equation}
 \hat{x}_t = x_t + s \nabla_{x_t} F(x_t, y)
\end{equation}
with factor $s$ to control the guidance scale.

\begin{table*}[ht]
\centering
\resizebox{2.1\columnwidth}{!}{
\begin{tabular}{c||ccccccc}
\toprule[1.2pt]
     \textbf{Dataset} &\textbf{Model}  & \textbf{Task} & FAD$_{\mathrm{VGG}} (\downarrow)$ & Instruction Acc. $(\uparrow)$ & Chord Rec. Acc. $(\uparrow)$ & Pitch His. $(\uparrow)$ & IO Interval $(\uparrow)$ \\ \midrule
     \multirow{11}*{In-house} & \multirow{5}*{AUDIT} & Extract & 1.67&0.39  &0.82 &0.62&0.54 \\
     &   & Remove  & 1.73& 0.65&0.86 &0.64&0.53 \\
    & & Add & 1.25 & 0.73 & 0.72& 0.64 & 0.54\\ 
     & &  Replace &  1.50& 0.62 & 0.83&0.63 &0.51 \\ 
    &  & Avg. & 1.54 &0.60 &0.81 &0.63 &0.53 \\ \cmidrule{2-8}
     &  \multirow{5}*{InstructME}  & Extract & 1.54& 0.56&0.86&0.69&0.68 \\
     &   & Remove   & 1.68 & 0.80& 0.88& 0.72&0.66 \\
    & & Add & 1.22 & 0.73&0.75 & 0.72&0.66  \\ 
     & &  Replace & 1.39 &0.62  &0.86&0.71&0.67 \\ 
    &  & Avg. &  \textbf{1.45}& \textbf{0.68}& \textbf{0.84}& \textbf{0.71}&\textbf{0.67} \\ %\cmidrule{2-8}
    % &  \multirow{1}*{AUDIT} & \multirow{2}*{Remix} &  0.49& 0.69 & 0.59&0.60 &0.46 \\ 
    % &  \multirow{1}*{InstructME} &  &  \textbf{0.45}& \textbf{0.73} & \textbf{0.70}&\textbf{0.72} &\textbf{0.64} \\
\midrule[1.0pt]
     \multirow{11}*{Slakh} & \multirow{5}*{AUDIT} & Extract & 4.91& 0.52 &0.66& 0.62&0.52 \\
     &   & Remove  &1.92 &0.57&0.57 &0.61 &0.51 \\
    & & Add & 3.11&0.87& 0.58& 0.63&0.47 \\ 
     & &  Replace & 4.08 &0.78& 0.55 & 0.62& 0.47\\ 
    &  & Avg. & \textbf{3.50} & 0.68& 0.59& 0.62&0.49 \\ \cmidrule{2-8}
     &  \multirow{5}*{InstructME}  & Extract & 5.04&  0.66& 0.71&0.70&0.71 \\
     &   & Remove  & 1.87 &0.79 &0.65 &0.70&0.69 \\
    & & Add & 3.15 & 0.87&0.66 &0.74&0.67  \\ 
     & &  Replace & 3.97 & 0.83 &0.65&0.74&0.69\\ 
    &  & Avg. & \textbf{3.50} & \textbf{0.79}& \textbf{0.67}& \textbf{0.72}&\textbf{0.69} \\ \bottomrule[1.2pt]
\end{tabular}
}
\caption{\textbf{Objective} Evaluation Results of different edit tasks on In-house and Slakh datasets. Avg. is the average result of several edit tasks including extract, remove, add and replace. FAD reflects the music quality, Instruction Acc., and Chord Rec. Acc., pitch His. and IO Interval can measure the harmony of edited music.}
\label{tab:objective}
    \vspace{-0.2cm}
\end{table*}

% \begin{table}[ht]
%     \centering
%     % \resizebox{1.0\columnwidth}{!}{
%     \begin{tabular}{c|ccccc}
%     \toprule
%     Model  & FAD & IA & CRA & PCH & IOI \\ \midrule
%     AUDIT & 0.49 & 0.69 & 0.59 & 0.60 & 0.46  \\
%     InstructME & 0.45 &  0.73 & 0.70 & 0.72 & 0.64  \\
%     \bottomrule
%     \end{tabular}
%     % }
%     \caption{Objective evaluation results of multi-round editing.}
%     \label{tab:remix}
% \end{table}
% \begin{table}[ht]
%     \centering
%     % \resizebox{1.0\columnwidth}{!}{
%     \begin{tabular}{lcc}
%     \toprule[1pt]
%     {Metric} & AUDIT & InstructME \\ \midrule
%     FAD$(\downarrow)$ & 0.49 & \textbf{0.45}  \\ 
%     % \midrule
%     IA\ \ $(\uparrow)$ & 0.69 & \textbf{0.73}   \\ 
%     % \midrule
%     CRA$(\uparrow)$ & 0.59  & \textbf{0.70}  \\
%     PCH$(\uparrow)$ & 0.60 & \textbf{0.72} \\
%     IOI\ $(\uparrow)$ & 0.46 & \textbf{0.64} \\
%     \bottomrule[1.2pt]
%     \end{tabular}
%     % }
%     \caption{Objective result of remix}
%     \label{tab:remix}
% \end{table}

\section{Experiments Setup}
\label{sec:dataset}
\subsection{Dataset}

% \subsubsection{Music Data} 
We collected 417 hours of music audio. Each audio file consists of multiple instrumental tracks. We resampled audios to 24khz sample rate and divided them into non-overlapping 10-second clips. % These audio samples are referred to as the 'in-house data'.
% For data cleaning: 
% \begin{itemize}
%     \item Normalize and unify the name of instruments.
%     \item For each clip, combine the tracks with the same instrument 
%     \item An energy-based Voice Activity Detection (VAD) is used to remove the clips with the most silence. 
% \end{itemize}

% \subsubsection{Triplet Data}
%For every audio clip, we sample pairs of versions that consist of different sets of instruments, and create a text instruction based on the difference of instruments. 
For each audio clip, we select pairs of versions with varying instrument compositions and generate a text instruction based on the instrument differences.
We use the clips generated before to prepare the triplet data $<$text instruction, source music, target music$>$ including remixing (1 Million), adding (0.3M) and replacement (0.3M), extracting (0.2M) and removing (0.2M) respectively. These music triplet data are referred to as the 'in-house data'. We show our detailed data processing methods in Appendix.

\subsubsection{Evaluation Data}
We evaluate the models on both in-domain data and out-domain data.
\begin{itemize}
    \item In-domain data: We split the in-house data randomly into two parts and use one subset to generate triplet data for evaluating the models.
    \item Out-domain data: To demonstrate the robustness of the system, we also evaluate the models on the Synthesized Lakh (Slakh) Dataset~\cite{manilow2019cutting} which is a dataset of multi-track audio and has no overlap with the training data.
\end{itemize}

\subsection{Evaluation Metric}
%所以我们分别从音频质量、文本正确率和音乐和谐性三个方面来进行测量。其中分别使用基于pann和vgg的fad算法来测量生成音频的质量，并使用xx文章提到的模型来测量命令正确率，同时受到xxx文章的启发，引入Chord acc，PCH_OA，IOI_OA来侧面刻画音乐和谐性
% Evaluating music edit models is challenging. So we conduct both subjective and objective evaluations to measure the performance of the proposed InstructME framework. 

% \subsubsection{Objective Evaluation}
% Yet music, which is sounds that are artificially organized in relation to the sensational moments, i.e. pitch (tone name), intensity (dynamics), rhythm (onset and offset), and timbre (instrumentation) ~\cite{weihs2016music}, cannot be edited without considering the complex interrelationship between sounds from different sources.
Music is sounds that are artificially organized in relation to the sensational moments, with complex interplay and multi-layered perceptual impact between pitch, intensity, rhythm and timbre.
Defining a single suitable metric to fully evaluate music is challenging, ~\cite{agostinelli2023musiclm, huang2023noise2music} focus evaluation on signal quality and semantics, whereas \cite{lv2023getmusic, ren2020popmag, yang2020evaluation} propose more direct evaluation approach based on musicality indicators, in order to achieve a more comprehensive evaluation of music, we proposed the following metrics to objectively evaluate the performance of edited music in three aspects:

\textbf{Music Quality}. We use the fr\'echet audio distance (FAD)\footnote{https://github.com/gudgud96/frechet-audio-distance}\cite{kilgour2019frechet} to measure the quality between edited music and target music, the audio classification model is implemented with 
% PANNs\cite{kong2020panns} and 
VGGish\cite{hershey2017cnn}. 

\textbf{Text Relevance}. We define the instruction accuracy(IA) metric to indicate the relevance of the text-music pair, the proposed editing tasks are all related to music tags such as instrument, mood and genre, so we calculate instruction accuracy according to the edited music tags and input command while tags are recognized with tagging models which implemented with \cite{lu2021spectnt}. 

\textbf{Harmony}. We introduce three metrics for quantitative evaluation:
\begin{itemize}
\item \textit{Chord Recognition Accuracy 
(CRA)}. 
%\XC{how to get chord model or label? }
Chord Recognition Accuracy measures the harmony coherence between edited music and target music. We acquire the chord progression sequences of both source and target music in the initial step, while the chord progression recognition model is implemented by ~\cite{cheuk2022jointist}. Then the alignment of these sequences is computed to determine the chord recognition accuracy.
% The chord accuracy is defined as:
% \begin{equation}
% CA = \frac{1}{N_{t} * N_{c}}\sum_{i=1}^{N_{t}}\sum_{j=1}^{N_{c}} I\left\{C_{i,j} = C_{i,j}^{'}\right\}
% \end{equation}
% where $N_{t}$ is the number of tracks, $N_{c}$ is the number of chords, $C_{i,j}$ and $C_{i,j}^{'}$ are the $j$-th chord in the $i$-th target track and edited track, respectively.
\item \textit{Pitch Class Histogram (PCH)}\footnote{https://github.com/RichardYang40148/mgeval\label{web}}. The pitch class histogram is a pitch content representation that is octave-independent. It uses 12 classes to represent the chromatic scale, which spans from 4 to 40. We calculate the distribution of pitches classes according to this histogram.
\item \textit{Inter-Onset Interval (IOI)}\textsuperscript{\ref {web}}. Inter-onset interval refers to the time between two note onsets within a bar. In our case, We quantize the intervals into 32 classes and calculate the distribution of interval classes. Regarding PCH and IOI, we further compute the averaging Overlapped Area of their distributions to quantize the musical harmony in terms of pitch and onset aspects. %the OA definition like:($\mathcal{D}_{OA}$)
% \begin{equation}
% \mathcal{D}_{OA}^{f} = \frac{1}{N_{t} * N_{b}}\sum_{i=1}^{N_{t}}\sum_{j=1}^{N_{b}} OA(\mathcal{P}_{i,j}^{f}, \hat{\mathcal{P}}_{i,j}^{f})
% \end{equation}
% where $f\in\left\{PCH, IOI\right\}$, $N_{t}$ is the number of tracks, $N_{b}$ is the number of bars, OA refers to the overlapping area of two distributions, $\mathcal{P}_{i,j}^{f}$ and $\hat{\mathcal{P}}_{i,j}^{f}$ are the distribution features in $j$-th bar and $i$-th track of edited music and target music. 
\end{itemize}
% More implementation details and calculation rules are shown in Appendix.

For objective evaluation, we generate 800 triplet data for each music editing task and evaluate them with these objective metrics.

% %在主观测试中，我们从三个维度来衡量目标系统的表现，
% \subsubsection{Subjective Evaluation} 
% In subjective evaluation, we also conduct the evaluation in the same three aspects. In order to better quantify scoring, we select ten samples randomly from test set for each music editing task, and we have designed 5 levels of evaluation criteria as reference for raters. 

% \subsection{Model Configuration}
% All diffusion models in the experiments are trained using 16 NVIDIA TESLA V100 32GB GPUs with a batch size of 1 per GPU for 200k steps. We optimize them with the AdamW optimizer. More details are shown in Appendix.

% warm up the learning rate for the first updates to a peak of $1e$-$4$, and then linear decay it.

% \subsection{Model Comparison} 
% Due to the difficulty in decoupling musical instruments, some methods~\cite{couairon2022diffedit,wang2023instructedit,huang2023collaborative} firstly generate the mask of areas to be edited and then restore it following the command is difficult to be applied in music editing. 
% Other methods~\cite{meng2021sdedit,liu2023audioldm} achieve the editing goal by adding noise to source music and then denoising the music with the command. However, all these methods rely heavily on a powerful diffusion model with a large amount of data pre-training (eg: 340k hours music data in Noise2Music~\cite{huang2023noise2music}), which is hard to train for us. 
% As a result, we use AUDIT~\cite{wang2023audit} as our baseline system in the following experiments.

%音乐和谐感的复杂定义使得寻找到一个直接的metric来测量音乐生成系统的表现变得困难，而受到xx的启发，我们尝试使用来侧面刻画

\section{Results and Analysis}

\subsection{Objective Evaluation Results}\label{subsec:obj}

% 3. 为了说明我们系统的鲁棒性和泛化性，除了和训练集同domain的in-house data，我们还提供了out-of-domain的slakh的评测结果。InstructME依然能比AUDIT取得更好的结果。

We compare against AUDIT~\cite{wang2023audit} trained on the same data and in the same VAE latent space, as our baseline system in all experiments.
As shown in Table~\ref{tab:objective}, InstructME outperforms AUDIT in terms of music quality, text relevance and harmony.
% gives an overview of the objective evaluation results of different tasks on In-house and Slakh datasets.  
% 1. 在不同类型的任务上，包括有确定答案的Extract和Remove操作和创造性的操作，InstructME都比AUDIT要好。
Specifically, operations such as extract and remove are tasks with definite answers, and these tasks emphasize the precision of generation. Alternatively, operations such as add, replace, and remix are tasks with indeterminate answers, and these tasks are more creatively oriented. These two types of tasks require the model to achieve stable and diverse output based on an accurate understanding of textual instructions. From Table~\ref{tab:objective}, it is observed that InstructME improves musical quality by 5.84\%, text relevance by 13.33\% and harmony up to 26.42\% compared to AUDIT on In-house dataset. These results demonstrate that our approach provides rich generated content in addition to capturing the difference between textual instructions.

% By comparing the results on different tasks, we found that our model can achieve better results than AUDIT on both tasks with deterministic answers(i.e. extract and remove) and creative tasks without deterministic answers (i.e. add, replace and Remix).
% 2. 我们在表格里面提供了三个维度的评测，文本相似度上达到comparable，在音乐相关的音质和和谐度的指标上，我们都有明显的提升。
% Three aspects of evaluation are shown in the Table including music quality, text relevance and harmony. Compared with baseline system AUDIT, InstructME achieves better performance on FAD, CRA, PCH and IOI metrics, and comparable results on text relevance, which demonstrates that InstructME offer higher quality and more harmony edit results and better melody consistency.
% 3. 为了说明我们系统的鲁棒性和泛化性，除了和训练集同domain的in-house data，我们还提供了out-of-domain的slakh的评测结果。InstructME依然能比AUDIT取得更好的结果。
% To measure the robustness and generalization,
% % in addition to the in-house data from the same domain, 
% we also provide evaluation results on out-of-domain dataset Slakh~\cite{manilow2019cutting} in Table~\ref{tab:objective}. InstructME maintains the same conclusion. %can still obtain better results than AUDIT.

\begin{table}[ht]
    \centering
    % \resizebox{1.0\columnwidth}{!}{
    \begin{tabular}{c|ccccc}
    \toprule
    Model  & FAD & IA & CRA & PCH & IOI \\ \midrule
    AUDIT & 0.49 & 0.69 & 0.59 & 0.60 & 0.46  \\
    InstructME & \textbf{0.45} &  \textbf{0.73} & \textbf{0.70} & \textbf{0.72} & \textbf{0.64}  \\
    \bottomrule
    \end{tabular}
    % }
    \caption{Objective evaluation results of remixing on In-house dataset.}
    \label{tab:remix}
\end{table}

To study the generalization ability of InstructME, we also test it on the public available dataset Slakh~\cite{manilow2019cutting} which is more challenging in maintaining harmony by including unseen chord progressions. Table~\ref{tab:objective} shows that InstructME achieves comparable performance with AUDIT on musical quality but better results on text relevance and harmony. Particularly for chord recognition accuracy, our method surpasses the previous method by 13.56\%.

As a novel and unique task proposed in this paper, we also evaluated the performance of InstructME and AUDIT on remix operation. As Table~\ref{tab:remix} indicates, compared to AUDIT, InstructME achieves a significant improvement in harmony related metrics(CRA by 18.6\%, PCH by 20\% and IOI by 39\%), and also achieves better results in music quality and text relevance.

% 为了评估我们提出的
% We report the detailed ablation study results in Table~\ref{tab:ablation} including the objective evaluation results of InstructME without chord condition and multi-scale aggregation. 

% In addition, we perform detailed ablation experiments to study the impact of the chord condition and Multi-Scale Aggregation proposed in this paper. The objective evaluation results of training InstructME without chord Condition and Multi-Scale Aggregation are listed in Table~\ref{tab:ablation} respectively.
% % It can be clearly observed that after the model removes the chord condition, 
% Without chord condition, three harmony related metrics including CRA, PCH and IOI degrad significantly which demonstrates that chord condition mechanism plays a crucial role in maintaining the harmony of music editing. After disabling MMulti-Scale Aggregation in U-Net, FAD gets worse noticeably, which shows Multi-Scale Aggregation ensure music audio quality when editing musics.

\begin{figure*}[ht!]
    \centering
    \includegraphics[width=2.1\columnwidth]{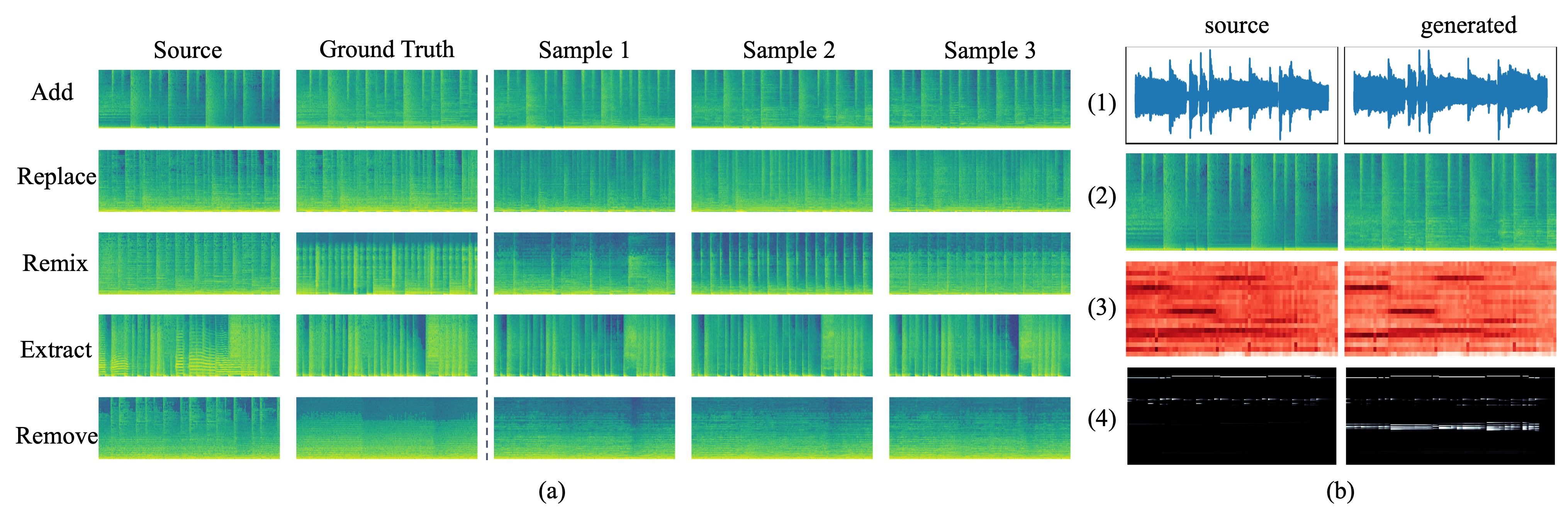}
    \caption{(a): Visualization of different editing tasks with three samples. All music segments are shown by spectrograms. (b): Comparison between source and edited music from four perspectives: (1) waveform (2) spectrogram (3) chord matrix (4) pitch matrix. The instruction command in the example is ``\textit{add acoustic guitar}''.}
    \label{fig:stable_diverse}
    % \vspace{-0.2cm}
\end{figure*}

\subsection{Subjective Evaluation Results}

\begin{table}[h!]
    \centering
    \resizebox{1.0\columnwidth}{!}{
    \begin{tabular}{cc|ccc}
    \toprule[1.2pt]
       Model & Length &  Quality &  Relevance & Harmony \\ \midrule
       \multirow{3}*{AUDIT} & 10s  & 2.79&2.94&3.01  \\
       & 30s &2.33&2.23&2.19 \\ 
       & 60s  &2.24&2.09&2.05 \\ \midrule
       \multirow{3}*{InstructME} & 10s  & 3.35&3.59&3.54 \\ 
       & 30s & 2.62&3.05&2.63 \\
       & 60s  &2.62&2.93&2.48  \\ \bottomrule[1.2pt]
    \end{tabular}
    }
    \caption{Subjective results of editing on music of different duration.}
    \label{tab:long}
\end{table}

% \begin{table}[ht]
%     \centering
%     \resizebox{1.00\columnwidth}{!}{
%     \begin{tabular}{ccccc}
%     \toprule[1.2pt]
%     Model & Task & Quality & Relevance & Harmony \\ \midrule
%     \multirow{6}*{AUDIT} &  Extract &2.74&2.67&3.22 \\
%     & Remove &3.03&3.23&3.21 \\
%     & Add &2.91&3.36&2.82 \\
%     & Replace &2.48&2.47&2.75 \\
%     & Remix &2.82&2.96&3.05 \\ 
%     % \cmidrule{2-5}
%     & Avg. &2.79&2.94&3.01 \\\midrule
%     \multirow{6}*{InstructME} & Extract &3.15&3.37&3.31 \\
%     & Remove &3.43&3.72&3.27 \\
%     & Add &3.53&3.76&3.73 \\ 
%     & Replace &3.22&3.36&3.44 \\
%     & Remix &3.44&3.72&3.93 \\ 
%     % \cmidrule{2-5}
%     & Avg. &\textbf{3.35}&\textbf{3.59}&\textbf{3.54} \\
%          \bottomrule[1.2pt]
%     \end{tabular}
%     }
%     \caption{\textbf{Subjective} Evaluation Results of different edit tasks on In-house dataset. Avg. is the average result of several edit tasks including extract, remove, add, replace and remix.}
%     % from three aspects including quality, relevance and harmony.}
%     \label{tab:subjective}
% \end{table}

We also conduct a subjective evaluation by outsourcing 10 testing samples(10s clip) per editing task for each labor. Mean Opinion Score(MOS) of scale 5 is used to compare the music quality, text relevance and harmony of two methods. To be more representative of real-world application scenarios, we also respectively generate 30-second and 60-second audio results, leveraging the chunk transformer's insensitivity to output length. A subjective evaluation of these generated long-duration music was also performed. We wrap all results in Table~\ref{tab:long} and the MOS scores show the superiority of our method over the baseline.

% In order to evaluate the performance of the InstructME in long-duration music editing tasks, we conduct the evaluation on music segments with 10s, 30s, and 60s respectively and list the results in Table~\ref{tab:long}. By comparing the editing results of different lengths, we find a significant degradation in the performance of both InstructME and AUDIT. However, with the help of chunk transformer, our InstructME can alleviate the negative impact of this music duration mismatch. 

% For a comprehensive evaluation of InstructME, in addition to objective testing, we present subjective evaluation results of different editing tasks in Table~\ref{tab:subjective}. InstructME clearly outperforms the baseline method in Music Quality, Text Relevance and harmony across all five edit tasks. By analyzing different tasks, we find that our method achieves more significant improvements on creative and generative tasks including adding, replacement and remixing. This also indicates that our model is able to encode music structure. Additionally, for manipulations with standard answers such as removing and extraction, our model still exhibits excellent performance compared to the baseline system.  

\subsection{Case Study}

\subsubsection{Chord Condition and Multi-Scale Aggregation}
We perform ablation experiments to study the impact of the chord condition and multi-scale aggregation. The objective evaluation results of training InstructME without chord condition and multi-scale aggregation are listed in Table~\ref{tab:ablation} respectively.
% It can be clearly observed that after the model removes the chord condition, 
The absence of chord conditioning is demonstrated to lead to a deterioration in harmony-related metrics such as CRA, PCH, and IOI. This observation underscores the critical role of the chord conditioning mechanism in holding the harmonicity of music editing. Moreover, the notable decline in FAD, subsequent to the deactivation of Multi-Scale Aggregation within the U-Net architecture, indicates its significant contribution to preserving the audio quality of music during the editing process.

% which demonstrates that chord condition mechanism plays a crucial role in maintaining the harmony of music editing. After disabling Multi-Scale Aggregation in U-Net, FAD gets worse noticeably, which shows Multi-Scale Aggregation ensure music audio quality when editing musics.

\begin{table}[ht]
    \centering
    \resizebox{1.0\columnwidth}{!}{
    \begin{tabular}{lccc}
    \toprule[1.2pt]
    {Metric} & InstructME & w/o Chord Condition & w/o MSA \\ \midrule
    FAD$(\downarrow)$ & 1.45 & 1.46$(\uparrow 0.01)$ &1.53$(\uparrow 0.08)$ \\ 
    % \midrule
    IA\ \ $(\uparrow)$ & 0.68 & 0.63$(\downarrow 0.05)$ & 0.66$(\downarrow 0.02)$  \\ 
    % \midrule
    CRA$(\uparrow)$ & 0.84  & 0.81$(\downarrow 0.03)$ &0.83$(\downarrow 0.01)$ \\
    PCH$(\uparrow)$ & 0.71 & 0.64$(\downarrow 0.07)$ & 0.72$(\uparrow 0.01)$\\
    IOI\ $(\uparrow)$ & 0.67 & 0.53$(\downarrow 0.14)$ & 0.67$(\downarrow 0.00)$\\
    \bottomrule[1.2pt]
    \end{tabular}
    }
    \caption{Impact of chord condition and multi-scale aggregation strategies. MSA denotes multi-scale aggregation here. Mean value over all edit operations for each metric is provided.}
    \label{tab:ablation}
    \vspace{-0.2cm}
\end{table}

\subsubsection{Stability and Diversity} 

In order to investigate the stability and diversity of generation within the InstructME framework, as depicted in Figure~\ref{fig:stable_diverse}(a), we employ a spectral diagram to visually represent each editing task. Each row contains the source music, the ground truth music and three samples generated by our method. As mentioned in Section~\ref{subsec:obj}, different editing operations require different modeling capabilities. For example, the remix demands creativity because of different interpretations of the same sound source. However, the remove manipulation requires accuracy since the model should accurately recognize instruments mentioned in the textual instructions. For tasks requiring precision, our model consistently generates results congruent with the ground truth. For creativity-oriented tasks, the visual representation illustrates the diversity present in the spectrogram of the sampled music segments, underscoring the capacity of InstructME to conceive and construe novel compositions derived from existing audio sources.

% To study the stability and diversity of InstructME, as shown in Figure~\ref{fig:stable_diverse}(a), we visualize each editing task in the form of a spectrum diagram. 
% It should be noted that all tasks are sampled thrice for three samples.
% to study the stability and diversity of InstructME. 
% Here, we divided the task into two categories to explore. For creative tasks such as adding, replacement and remixing tasks, there is no definitive answer because musical instruments have multiple ways of interpretation on the premise of ensuring harmony. 
% From the figure, it can be seen that the music segments we sampled are diverse on the spectrogram, indicating that our InstructME has the ability to create and interpret new works based on existing audio.
% For manipulations including extraction and removing, our model can stably generate consistent results with ground truth although sampled several times randomly with different seeds. 

% \begin{table}[h!]
%     \centering
%     \resizebox{1.0\columnwidth}{!}{
%     \begin{tabular}{cc|ccc}
%     \toprule
%        Length & Model &  Quality &  Relevance & Harmony \\ \midrule
%        \multirow{2}*{10s} & AUDIT  & 2.79&2.94&3.01  \\
%        & InstructME  &3.35&3.59&3.54  \\ \midrule
%        \multirow{2}*{30s} & AUDIT  &2.33&2.23&2.19  \\
%        & InstructME  &2.62&3.05&2.63  \\ \midrule
%        \multirow{2}*{60s} & AUDIT  &2.24&2.09&2.05  \\
%        & InstructME  &2.62&2.93&2.48  \\ \bottomrule
%     \end{tabular}
%     }
%     \caption{Results of long music editing with subjective evaluation metrics.}
%     \label{tab:long}
% \end{table}

\begin{figure}[h]
    \centering
    \includegraphics[width=1.0\columnwidth]{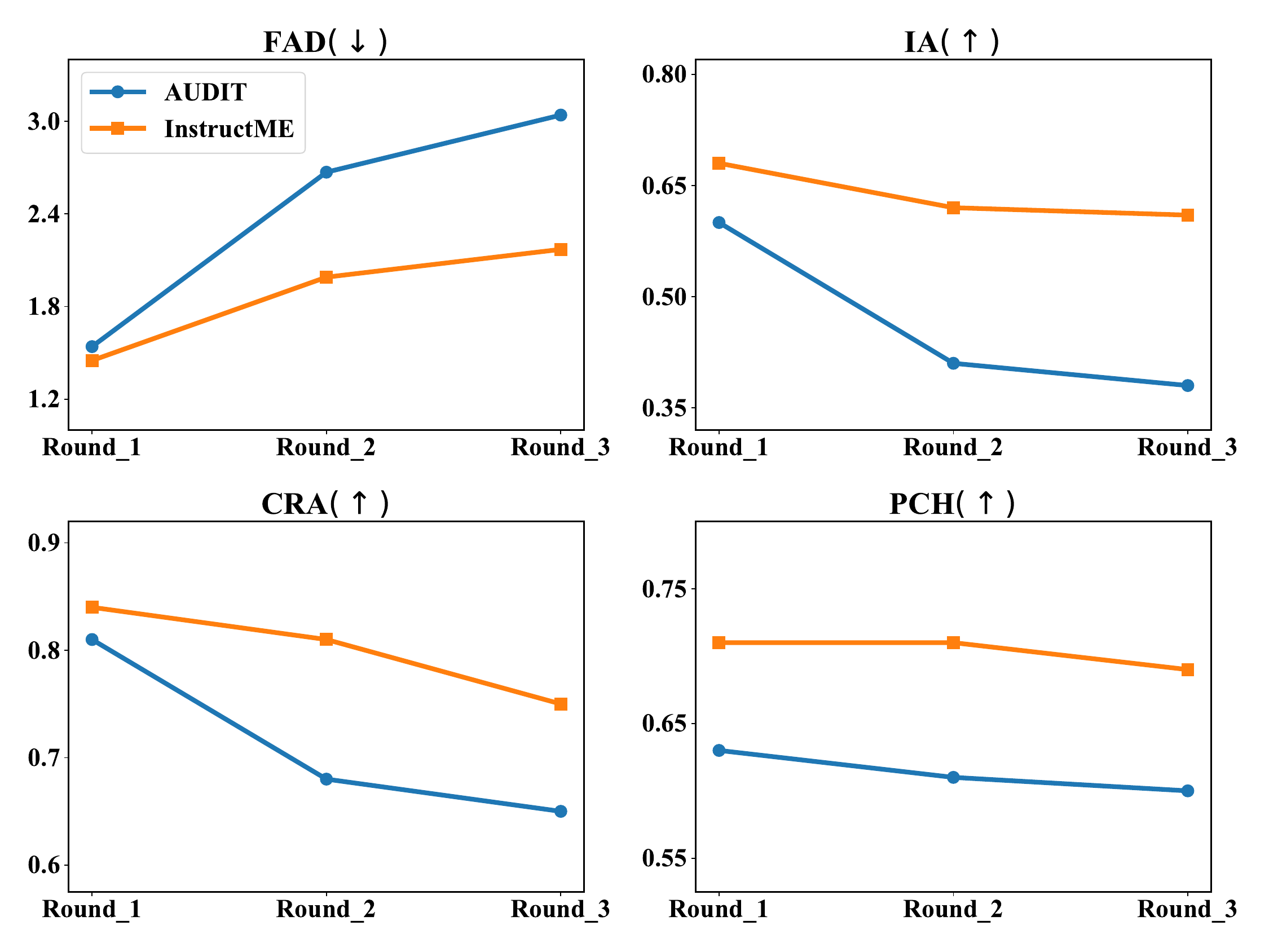}
    \caption{Line chart of objective evaluation results of three-round editing. Compared to AUDIT, InstructME shows a smaller decrease in metrics related to harmony and text relevance, while also exhibiting a smaller degradation in music quality(FAD).}
    \label{fig:trend}
    % \vspace{-0.3cm}
\end{figure}

\subsubsection{Consistency and Harmony}

% \begin{figure}
%     \centering
%     \includegraphics[width=0.9\columnwidth]{CameraReady/figure/consist.png}
%     \caption{\textbf{Visualization} of comparison between source and edited music with different figures including (a) waveform (b) spectrogram (c) chord matrix (d) pitch matrix. It's a case study for the adding task with instruction command ``\textit{add acoustic guitar}''.}
%     \label{fig:consistency}
% \end{figure}
In Figure 4(b), we present an illustrative study to elucidate aspects of consistency and harmony. In the example, we take ``\textit{Add acoustic guitar}'' as the textual instruction. The waveforms in Figure 4(b)(1) indicate that the beat timings before and after editing are meticulously synchronized, which demonstrates InstructME's capacity to maintain temporal consistency through the editing process. The temporal consistency is also observed in the corresponding spectrograms in Figure 4(b)(2), which exhibit energy spikes at identical times. Subsequently, we proceed to extract and compare the chord progression matrices from the source and generated musical segments as portrayed in Figure 4(b)(3). The intensity of the color is indicative of the predicted chord probability. Upon scrutinizing the probability patterns between the source and generated music, it is concluded that our method is able to preserve the musical harmony of the source. The last row in Figure~\ref{fig:stable_diverse}(b) is the pitch matrix. In this representation, the uppermost pair of white stripes corresponds to the pitch pertaining to the piano and drum, respectively. Remarkably, the mere variations of these two lines between source and generated music are evidence of the consistency maintenance. The third stripe corresponds to the guitar's pitch information. Furthermore, the absence of other instruments from the generated music validates our model's precise and controlled behavior.

In pursuit of a more comprehensive understanding of our model's persistence in consistency and harmony, we undertake a series of multi-round editing experiments, comparing the musical outcomes of InstructME and AUDIT following repeated editing iterations. Results are listed in Figure~\ref{fig:trend}. Notably, both our method and the baseline approaches display a gradual diminishment in performance as the iterative editing processes unfold. However, InstructME exhibits a comparatively marginal deterioration, particularly in terms of music harmony. Importantly, the remaining metrics remain well within an acceptable range, attesting to the robustness of the model's performance.

% \subsection{Long-Duration Music Edit}
% 为什么想要long music edit，（1）直接跑会有mismatch，（2）分别跑，会有mismatch

% \begin{table}[h]
%     \centering
%     \begin{tabular}{c|cc}
%     \toprule
%      Model  &  Memory & Speed \\ \midrule
%     Transformer   &  & \\
%     Chunk Transformer & & \\
%     \bottomrule
%     \end{tabular}
%     \caption{The GPU memory usage and speed of U-Net with or without chunk transformer during sampling process.}
%     \label{tab:long_memory}
% \end{table}

% \section{Limitation}

\section{Conclusion}
In this work, we introduce InstructME, a music editing and remixing framework based on latent diffusion models. 
For InstructME, we enhance the U-Net with multi-scale aggregation and chord condition to improve the harmony and consistency of edited music, and introduce chunk transformer to extend the long-term music generation capabilities.
% We propose multi-scale aggregation and chord condition to improve the harmony and consistency of edited music, and introduce chunk transformer to extend the long-term music generation capabilities of InstructME.
% We train InstructME using our generated 2M triplet music data pair, which shows that our InstructME can effectively edit source music based on simple editing instructions, while preserving certain musical components and generating harmonious results that align with the semantic information conveyed in the instructions. 
To evaluate the efficacy of music editing results, we establish several quantitative metrics and conduct experimental trials to validate them. Our findings indicate that the proposed InstructME outperforms the baselines in both subjective and objective experiments, which shows that our InstructME can effectively edit source music based on simple editing instructions, while preserving certain musical components and generating harmonious results that align with the semantic information conveyed in the instructions. 
In future endeavors, we aim to expand the scope and practical value of InstructME by exploring more intricate music editing tasks, such as structured editing.

\bibliography{aaai24}

\begin{thebibliography}{43}
\providecommand{\natexlab}[1]{#1}

\bibitem[{Agostinelli et~al.(2023)Agostinelli, Denk, Borsos, Engel, Verzetti,
  Caillon, Huang, Jansen, Roberts, Tagliasacchi
  et~al.}]{agostinelli2023musiclm}
Agostinelli, A.; Denk, T.~I.; Borsos, Z.; Engel, J.; Verzetti, M.; Caillon, A.;
  Huang, Q.; Jansen, A.; Roberts, A.; Tagliasacchi, M.; et~al. 2023.
\newblock Musiclm: Generating music from text.
\newblock \emph{arXiv preprint arXiv:2301.11325}.

\bibitem[{Cheuk et~al.(2022)Cheuk, Choi, Kong, Li, Won, Hung, Wang, and
  Herremans}]{cheuk2022jointist}
Cheuk, K.~W.; Choi, K.; Kong, Q.; Li, B.; Won, M.; Hung, A.; Wang, J.-C.; and
  Herremans, D. 2022.
\newblock Jointist: Joint learning for multi-instrument transcription and its
  applications.
\newblock \emph{arXiv preprint arXiv:2206.10805}.

\bibitem[{Copet et~al.(2023)Copet, Kreuk, Gat, Remez, Kant, Synnaeve, Adi, and
  D{\'e}fossez}]{copet2023simple}
Copet, J.; Kreuk, F.; Gat, I.; Remez, T.; Kant, D.; Synnaeve, G.; Adi, Y.; and
  D{\'e}fossez, A. 2023.
\newblock Simple and Controllable Music Generation.
\newblock \emph{arXiv preprint arXiv:2306.05284}.

\bibitem[{Dhariwal and Nichol(2021)}]{dhariwal2021diffusion}
Dhariwal, P.; and Nichol, A. 2021.
\newblock Diffusion models beat gans on image synthesis.
\newblock \emph{Advances in neural information processing systems}, 34:
  8780--8794.

\bibitem[{Donahue et~al.(2023)Donahue, Caillon, Roberts, Manilow, Esling,
  Agostinelli, Verzetti, Simon, Pietquin, Zeghidour
  et~al.}]{donahue2023singsong}
Donahue, C.; Caillon, A.; Roberts, A.; Manilow, E.; Esling, P.; Agostinelli,
  A.; Verzetti, M.; Simon, I.; Pietquin, O.; Zeghidour, N.; et~al. 2023.
\newblock SingSong: Generating musical accompaniments from singing.
\newblock \emph{arXiv preprint arXiv:2301.12662}.

\bibitem[{Fagerjord(2010)}]{fagerjord2010after}
Fagerjord, A. 2010.
\newblock After convergence: YouTube and remix culture.
\newblock \emph{International handbook of internet research}, 187--200.

\bibitem[{He et~al.(2016)He, Zhang, Ren, and Sun}]{he2016deep}
He, K.; Zhang, X.; Ren, S.; and Sun, J. 2016.
\newblock Deep residual learning for image recognition.
\newblock In \emph{Proceedings of the IEEE conference on computer vision and
  pattern recognition}, 770--778.

\bibitem[{Hershey et~al.(2017)Hershey, Chaudhuri, Ellis, Gemmeke, Jansen,
  Moore, Plakal, Platt, Saurous, Seybold et~al.}]{hershey2017cnn}
Hershey, S.; Chaudhuri, S.; Ellis, D.~P.; Gemmeke, J.~F.; Jansen, A.; Moore,
  R.~C.; Plakal, M.; Platt, D.; Saurous, R.~A.; Seybold, B.; et~al. 2017.
\newblock CNN architectures for large-scale audio classification.
\newblock In \emph{2017 ieee international conference on acoustics, speech and
  signal processing (icassp)}, 131--135. IEEE.

\bibitem[{Hertz et~al.(2022)Hertz, Mokady, Tenenbaum, Aberman, Pritch, and
  Cohen-Or}]{hertz2022prompt}
Hertz, A.; Mokady, R.; Tenenbaum, J.; Aberman, K.; Pritch, Y.; and Cohen-Or, D.
  2022.
\newblock Prompt-to-prompt image editing with cross attention control.
\newblock \emph{arXiv preprint arXiv:2208.01626}.

\bibitem[{Ho, Jain, and Abbeel(2020)}]{ho2020denoising}
Ho, J.; Jain, A.; and Abbeel, P. 2020.
\newblock Denoising diffusion probabilistic models.
\newblock \emph{Advances in neural information processing systems}, 33:
  6840--6851.

\bibitem[{Ho and Salimans(2022)}]{ho2022classifier}
Ho, J.; and Salimans, T. 2022.
\newblock Classifier-free diffusion guidance.
\newblock \emph{arXiv preprint arXiv:2207.12598}.

\bibitem[{Holz(2023)}]{midjourney2023}
Holz, D. 2023.
\newblock Midjourney. Artificial Intelligence platform. Accessible at
  https://www.midjourney.com.
\newblock \url{https://www.midjourney.com/}.
\newblock Accessed: 2023-07-31.

\bibitem[{Huang et~al.(2022)Huang, Jansen, Lee, Ganti, Li, and
  Ellis}]{huang2022mulan}
Huang, Q.; Jansen, A.; Lee, J.; Ganti, R.; Li, J.~Y.; and Ellis, D.~P. 2022.
\newblock Mulan: A joint embedding of music audio and natural language.
\newblock \emph{arXiv preprint arXiv:2208.12415}.

\bibitem[{Huang et~al.(2023{\natexlab{a}})Huang, Park, Wang, Denk, Ly, Chen,
  Zhang, Zhang, Yu, Frank et~al.}]{huang2023noise2music}
Huang, Q.; Park, D.~S.; Wang, T.; Denk, T.~I.; Ly, A.; Chen, N.; Zhang, Z.;
  Zhang, Z.; Yu, J.; Frank, C.; et~al. 2023{\natexlab{a}}.
\newblock Noise2music: Text-conditioned music generation with diffusion models.
\newblock \emph{arXiv preprint arXiv:2302.03917}.

\bibitem[{Huang et~al.(2023{\natexlab{b}})Huang, Huang, Yang, Ren, Liu, Li, Ye,
  Liu, Yin, and Zhao}]{huang2023make}
Huang, R.; Huang, J.; Yang, D.; Ren, Y.; Liu, L.; Li, M.; Ye, Z.; Liu, J.; Yin,
  X.; and Zhao, Z. 2023{\natexlab{b}}.
\newblock Make-an-audio: Text-to-audio generation with prompt-enhanced
  diffusion models.
\newblock \emph{arXiv preprint arXiv:2301.12661}.

\bibitem[{Jeong, Kwon, and Uh(2023)}]{jeong2023training}
Jeong, J.; Kwon, M.; and Uh, Y. 2023.
\newblock Training-free Style Transfer Emerges from h-space in Diffusion
  models.
\newblock \emph{arXiv preprint arXiv:2303.15403}.

\bibitem[{Karras et~al.(2020)Karras, Laine, Aittala, Hellsten, Lehtinen, and
  Aila}]{karras2020analyzing}
Karras, T.; Laine, S.; Aittala, M.; Hellsten, J.; Lehtinen, J.; and Aila, T.
  2020.
\newblock Analyzing and improving the image quality of stylegan.
\newblock In \emph{Proceedings of the IEEE/CVF conference on computer vision
  and pattern recognition}, 8110--8119.

\bibitem[{Kilgour et~al.(2019)Kilgour, Zuluaga, Roblek, and
  Sharifi}]{kilgour2019frechet}
Kilgour, K.; Zuluaga, M.; Roblek, D.; and Sharifi, M. 2019.
\newblock Fr{\'e}chet audio distance: A reference-free metric for evaluating
  music enhancement algorithms.
\newblock In \emph{INTERSPEECH}, 2350--2354.

\bibitem[{Kwon, Jeong, and Uh(2022)}]{kwon2022diffusion}
Kwon, M.; Jeong, J.; and Uh, Y. 2022.
\newblock Diffusion models already have a semantic latent space.
\newblock \emph{arXiv preprint arXiv:2210.10960}.

\bibitem[{Lam et~al.(2023)Lam, Tian, Li, Yin, Feng, Tu, Ji, Xia, Ma, Song
  et~al.}]{lam2023efficient}
Lam, M.~W.; Tian, Q.; Li, T.; Yin, Z.; Feng, S.; Tu, M.; Ji, Y.; Xia, R.; Ma,
  M.; Song, X.; et~al. 2023.
\newblock Efficient Neural Music Generation.
\newblock \emph{arXiv preprint arXiv:2305.15719}.

\bibitem[{Liu et~al.(2023{\natexlab{a}})Liu, Chen, Yuan, Mei, Liu, Mandic,
  Wang, and Plumbley}]{liu2023audioldm}
Liu, H.; Chen, Z.; Yuan, Y.; Mei, X.; Liu, X.; Mandic, D.; Wang, W.; and
  Plumbley, M.~D. 2023{\natexlab{a}}.
\newblock Audioldm: Text-to-audio generation with latent diffusion models.
\newblock \emph{arXiv preprint arXiv:2301.12503}.

\bibitem[{Liu et~al.(2023{\natexlab{b}})Liu, Park, Azadi, Zhang, Chopikyan, Hu,
  Shi, Rohrbach, and Darrell}]{liu2023more}
Liu, X.; Park, D.~H.; Azadi, S.; Zhang, G.; Chopikyan, A.; Hu, Y.; Shi, H.;
  Rohrbach, A.; and Darrell, T. 2023{\natexlab{b}}.
\newblock More control for free! image synthesis with semantic diffusion
  guidance.
\newblock In \emph{Proceedings of the IEEE/CVF Winter Conference on
  Applications of Computer Vision}, 289--299.

\bibitem[{Lu et~al.(2021)Lu, Wang, Won, Choi, and Song}]{lu2021spectnt}
Lu, W.-T.; Wang, J.-C.; Won, M.; Choi, K.; and Song, X. 2021.
\newblock SpecTNT: A time-frequency transformer for music audio.
\newblock \emph{arXiv preprint arXiv:2110.09127}.

\bibitem[{Lugmayr et~al.(2022)Lugmayr, Danelljan, Romero, Yu, Timofte, and
  Van~Gool}]{lugmayr2022repaint}
Lugmayr, A.; Danelljan, M.; Romero, A.; Yu, F.; Timofte, R.; and Van~Gool, L.
  2022.
\newblock Repaint: Inpainting using denoising diffusion probabilistic models.
\newblock In \emph{Proceedings of the IEEE/CVF Conference on Computer Vision
  and Pattern Recognition}, 11461--11471.

\bibitem[{Lv et~al.(2023)Lv, Tan, Lu, Ye, Zhang, Bian, and
  Yan}]{lv2023getmusic}
Lv, A.; Tan, X.; Lu, P.; Ye, W.; Zhang, S.; Bian, J.; and Yan, R. 2023.
\newblock GETMusic: Generating Any Music Tracks with a Unified Representation
  and Diffusion Framework.
\newblock \emph{arXiv preprint arXiv:2305.10841}.

\bibitem[{Manilow et~al.(2019)Manilow, Wichern, Seetharaman, and
  Le~Roux}]{manilow2019cutting}
Manilow, E.; Wichern, G.; Seetharaman, P.; and Le~Roux, J. 2019.
\newblock Cutting music source separation some Slakh: A dataset to study the
  impact of training data quality and quantity.
\newblock In \emph{2019 IEEE Workshop on Applications of Signal Processing to
  Audio and Acoustics (WASPAA)}, 45--49. IEEE.

\bibitem[{Meng et~al.(2021)Meng, He, Song, Song, Wu, Zhu, and
  Ermon}]{meng2021sdedit}
Meng, C.; He, Y.; Song, Y.; Song, J.; Wu, J.; Zhu, J.-Y.; and Ermon, S. 2021.
\newblock Sdedit: Guided image synthesis and editing with stochastic
  differential equations.
\newblock \emph{arXiv preprint arXiv:2108.01073}.

\bibitem[{Nichol et~al.(2021)Nichol, Dhariwal, Ramesh, Shyam, Mishkin, McGrew,
  Sutskever, and Chen}]{nichol2021glide}
Nichol, A.; Dhariwal, P.; Ramesh, A.; Shyam, P.; Mishkin, P.; McGrew, B.;
  Sutskever, I.; and Chen, M. 2021.
\newblock Glide: Towards photorealistic image generation and editing with
  text-guided diffusion models.
\newblock \emph{arXiv preprint arXiv:2112.10741}.

\bibitem[{Raffel et~al.(2020)Raffel, Shazeer, Roberts, Lee, Narang, Matena,
  Zhou, Li, and Liu}]{raffel2020exploring}
Raffel, C.; Shazeer, N.; Roberts, A.; Lee, K.; Narang, S.; Matena, M.; Zhou,
  Y.; Li, W.; and Liu, P.~J. 2020.
\newblock Exploring the limits of transfer learning with a unified text-to-text
  transformer.
\newblock \emph{The Journal of Machine Learning Research}, 21(1): 5485--5551.

\bibitem[{Ramesh et~al.(2022)Ramesh, Dhariwal, Nichol, Chu, and
  Chen}]{ramesh2022hierarchical}
Ramesh, A.; Dhariwal, P.; Nichol, A.; Chu, C.; and Chen, M. 2022.
\newblock Hierarchical text-conditional image generation with clip latents.
\newblock \emph{arXiv preprint arXiv:2204.06125}.

\bibitem[{Ren et~al.(2020)Ren, He, Tan, Qin, Zhao, and Liu}]{ren2020popmag}
Ren, Y.; He, J.; Tan, X.; Qin, T.; Zhao, Z.; and Liu, T.-Y. 2020.
\newblock Popmag: Pop music accompaniment generation.
\newblock In \emph{Proceedings of the 28th ACM international conference on
  multimedia}, 1198--1206.

\bibitem[{Rombach et~al.(2022)Rombach, Blattmann, Lorenz, Esser, and
  Ommer}]{rombach2022high}
Rombach, R.; Blattmann, A.; Lorenz, D.; Esser, P.; and Ommer, B. 2022.
\newblock High-resolution image synthesis with latent diffusion models.
\newblock In \emph{Proceedings of the IEEE/CVF conference on computer vision
  and pattern recognition}, 10684--10695.

\bibitem[{Ronneberger, Fischer, and Brox(2015)}]{ronneberger2015u}
Ronneberger, O.; Fischer, P.; and Brox, T. 2015.
\newblock U-net: Convolutional networks for biomedical image segmentation.
\newblock In \emph{Medical Image Computing and Computer-Assisted
  Intervention--MICCAI 2015: 18th International Conference, Munich, Germany,
  October 5-9, 2015, Proceedings, Part III 18}, 234--241. Springer.

\bibitem[{Saharia et~al.(2022)Saharia, Chan, Saxena, Li, Whang, Denton,
  Ghasemipour, Gontijo~Lopes, Karagol~Ayan, Salimans
  et~al.}]{saharia2022photorealistic}
Saharia, C.; Chan, W.; Saxena, S.; Li, L.; Whang, J.; Denton, E.~L.;
  Ghasemipour, K.; Gontijo~Lopes, R.; Karagol~Ayan, B.; Salimans, T.; et~al.
  2022.
\newblock Photorealistic text-to-image diffusion models with deep language
  understanding.
\newblock \emph{Advances in Neural Information Processing Systems}, 35:
  36479--36494.

\bibitem[{Schneider, Jin, and Sch{\"o}lkopf(2023)}]{schneider2023mo}
Schneider, F.; Jin, Z.; and Sch{\"o}lkopf, B. 2023.
\newblock Mo$\backslash$\^{} usai: Text-to-Music Generation with Long-Context
  Latent Diffusion.
\newblock \emph{arXiv preprint arXiv:2301.11757}.

\bibitem[{Song et~al.(2020)Song, Sohl-Dickstein, Kingma, Kumar, Ermon, and
  Poole}]{song2020score}
Song, Y.; Sohl-Dickstein, J.; Kingma, D.~P.; Kumar, A.; Ermon, S.; and Poole,
  B. 2020.
\newblock Score-based generative modeling through stochastic differential
  equations.
\newblock \emph{arXiv preprint arXiv:2011.13456}.

\bibitem[{Vaswani et~al.(2017)Vaswani, Shazeer, Parmar, Uszkoreit, Jones,
  Gomez, Kaiser, and Polosukhin}]{vaswani2017attention}
Vaswani, A.; Shazeer, N.; Parmar, N.; Uszkoreit, J.; Jones, L.; Gomez, A.~N.;
  Kaiser, {\L}.; and Polosukhin, I. 2017.
\newblock Attention is all you need.
\newblock \emph{Advances in neural information processing systems}, 30.

\bibitem[{Wang et~al.(2023)Wang, Ju, Tan, He, Wu, Bian, and
  Zhao}]{wang2023audit}
Wang, Y.; Ju, Z.; Tan, X.; He, L.; Wu, Z.; Bian, J.; and Zhao, S. 2023.
\newblock AUDIT: Audio Editing by Following Instructions with Latent Diffusion
  Models.
\newblock \emph{arXiv preprint arXiv:2304.00830}.

\bibitem[{Waysdorf(2021)}]{waysdorf2021remix}
Waysdorf, A.~S. 2021.
\newblock Remix in the age of ubiquitous remix.
\newblock \emph{Convergence}, 27(4): 1129--1144.

\bibitem[{Wierstorf et~al.(2017)Wierstorf, Ward, Mason, Grais, Hummersone, and
  Plumbley}]{wierstorf2017perceptual}
Wierstorf, H.; Ward, D.; Mason, R.; Grais, E.~M.; Hummersone, C.; and Plumbley,
  M.~D. 2017.
\newblock Perceptual evaluation of source separation for remixing music.
\newblock In \emph{Audio Engineering Society Convention 143}. Audio Engineering
  Society.

\bibitem[{Yang et~al.(2022)Yang, Firodiya, Bryan, and Kim}]{yang2022don}
Yang, H.; Firodiya, S.; Bryan, N.~J.; and Kim, M. 2022.
\newblock Don’t Separate, Learn To Remix: End-To-End Neural Remixing With
  Joint Optimization.
\newblock In \emph{ICASSP 2022-2022 IEEE International Conference on Acoustics,
  Speech and Signal Processing (ICASSP)}, 116--120. IEEE.

\bibitem[{Yang and Lerch(2020{\natexlab{a}})}]{yang2020evaluation}
Yang, L.-C.; and Lerch, A. 2020{\natexlab{a}}.
\newblock On the evaluation of generative models in music.
\newblock \emph{Neural Computing and Applications}, 32(9): 4773--4784.

\bibitem[{Yang and Lerch(2020{\natexlab{b}})}]{yang2020remixing}
Yang, L.-C.; and Lerch, A. 2020{\natexlab{b}}.
\newblock Remixing music with visual conditioning.
\newblock In \emph{2020 IEEE International Symposium on Multimedia (ISM)},
  181--188. IEEE.

\end{thebibliography}
\clearpage

\section{Appendix}

\subsection{Data}
\subsubsection{Data Preprocess}
For training the UniME, we collected 417 hours music audio files which contain different instrumental tracks, which is named in-house data. Then, we preprocess
them by resampling their sample rate to 24khz and splitting them into non-overlapping 10-second clips with sliding window. For data cleaning, 
\begin{itemize}
    \item Normalize and unify the name of instruments.
    \item For each clip, combine the tracks with the same instrument. Eg: combine \textit{acoustic guitar (left)} and \textit{acoustic guitar(right)} into \textit{acoustic guitar}.
    \item An energy-based Voice Activity Detection (VAD) is used to remove the clips with the most silence. 
\end{itemize}

\subsubsection{Triplet Data Generation}
Then we use the clips generated before to prepare the music-to-music triplet data including <instruction, source music, target music> for training the InstructME. As shown in Table~\ref{tab:command}, we only provide five different music edit tasks in this work. And the workflow of triplet generation is described in the following:
\begin{itemize}
    \item \textbf{ReMix}: Firstly, we select one clip randomly from the database, mix all the instrument tracks and regard the mixed clip as source music. Then we obtain the rhythm and time step information of the source music. In the following, we retrieve another music clip which has the same rhythm as target music and aligns the source-target music pair with time step information. Finally, we generate the text command as the template listed in Table~\ref{tab:command}.
    % Firstly, we select one clip randomly from the database and then mix all the instrument tracks by averaging stereo mixes to mono. It will be regarded as the source music. To find the paired target music, we obtain the chord and step information with chord and step recognition models. In the following, we retrieve another music clip with the same chord progression as the target music from the database and align them with step information. Finally, we generate the text command as the template listed in Table~\ref{tab:command}.
    \item \textbf{Add}: Similarly, we randomly choose one clip from the database in the first. And $i$ ($i\in [1,2,3,4]$) instrument tracks of the clips will be selected and mixed to get the source music. Then, we can get the target music by selecting another instrument track and combining it with the source music. 
    \item \textbf{Remove}: Removing task can be regarded as the reverse edit operation of adding, so we can generate the removing triplet data by reversing the pairs of adding data.
    \item \textbf{Extract}: For extracting task, we select one clip and choose one of the instrument tracks as target music. Then, we mix this track with some other instruments as the source music.
    \item \textbf{Replace}: To generate the replace data pairs, we choose two different instrument tracks and mix them with some other instruments. 
\end{itemize}

\subsubsection{Text Instruction Templates}

we provide five different music edit tasks in this work. And the workflow of triplet generation is described in Table~\ref{tab:command}, there are some specific examples of command:
\begin{itemize}
\item add distorted electric guitar
\item add synthesizer
\item add piccolo
\item extract viola
\item extract synth strings
\item extract string section
\item remove accordion
\item remove string section
\item remove clarinet
\item replace distorted electric guitar with electric guitar
\item replace guitar synth with synth pad
\item replace flute with accordion
\item remix with bass, piano, strings
\item remix with guitar, bass, drums
\item remix with drums, bass, guitar, piano
\item remix to Rock genre
\item remix to R\&B genre
\item remix to Jazz genre
\end{itemize}

\begin{table}[ht]
\centering
%\resizebox{.95\columnwidth}{!}{
\begin{tabular}{l|l}
\toprule
    Task & Text Command \\ \midrule
    Remix & Remix with \{instrument/genre\} \\
    Add & Add \{instrument\} \\
    Remove & Remove \{instrument\} \\
    Extract & Extract \{instrument\} \\
    Replace & Replace \{instrument A\} with \{instrument B\} \\ \bottomrule
\end{tabular}
\caption{The text command templates of each edit task.}
\label{tab:command}
\end{table}

\begin{table*}[ht]
\centering
\resizebox{2.1\columnwidth}{!}{
\begin{tabular}{c|p{0.2\columnwidth}|p{0.28\columnwidth}|p{0.28\columnwidth}|p{0.28\columnwidth}|p{0.28\columnwidth}|p{0.28\columnwidth}}
\toprule[1.2pt]
    Metrics & Task & 1 point & 2 points & 3 points& 4 points& 5 points\\ \midrule
    Music Quality  & & The sound quality is very poor, basically it's all noise. & Mostly noise, some instruments can be heard, some instruments sound normal. & There is noise, but it does not affect the performance of the instruments, and most of the instruments sound normal. & There is almost no noise, and the timbre is normal. & There is no noise at all, and the sound of the instrument is very pure, like a professional recording studio. \\ \midrule
    \multirow{2}{*}{Text Relevance}  & Editing for instruments & Nothing to do with the command, no instrument matches. & There are one or two instruments that match with command. & The more obvious instruments all match with command, but there are still at least two instruments that do not match. & Basically right, but missing one. & All correct. \\ \cmidrule{2-7}
    & Editing for mood or genre & It doesn't match with command, and it's obvious that it's not right. & Not right, but at least not the opposite. & There is no right or wrong, the genre or mood is rather vague. & The direction is obvious, and it can basically be judged to be correct. & Direction is clear and correct. \\ \midrule
    \multirow{2}{*}{Harmony}  & Atomic operation & It doesn't harmony with the source music, you can obviously hear it. & It's not very harmonious with the source music, but the onset is correct. & Although it doesn't harmonize with the source music, it doesn't affect hearing. & Basically the same as target. & Same or different from target, but the overall music is very nice and has a strong sense of harmony \\ \cmidrule{2-7}
    & Remix & It doesn't match with command, and it's obvious that it's not right. It's incongruous with the vocal, you can obviously hear it. & It doesn't coordinate well with the vocals, only the alignment of the beat can be heard. & The beats are aligned, but also basically in harmony with the vocals. & The beats and chords are basically aligned, and it sounds like a song. & It is spliced together in complete harmony with the vocals and has a reinterpretation part. \\
    \bottomrule[1.2pt]
\end{tabular}
}
\caption{Quantitative benchmark and description of difference tasks.}
\label{tab:benchmark}
\end{table*}

\subsection{Metric Calculation}

This section endeavors to expound upon the computation of evaluation metrics. As stated in the primary paper, a multitude of metrics have been chosen to evaluate edited music, primarily categorized into three dimensions: music quality, text relevance, and harmony. The corresponding metrics and computational methodologies are delineated as follows.
\begin{itemize}
\item \textit{Fr\'echet Audio Distance (FAD)}. FAD is defined as a measure of quality difference between a given audio and target audio (clean, high-quality audio). In contrast to extant audio evaluation metrics, FAD does not scrutinize individual audio clips. Rather, it compares embedding statistics derived from the complete evaluation set with those generated from a massive compilation of unaltered music. In general, FAD can be defined as:
\begin{equation}
F(\mathcal{N}_{t}, \mathcal{N}_{e}) = \left \| \mu_{t} - \mu_{e} \right\|^{2} + tr(\Sigma_{t} + \Sigma_{e} - 2\sqrt{\Sigma_{t}\Sigma_{e}})
\end{equation}
where $\mathcal{N}_{e}(\mu_{e}, \Sigma_{e})$ is the evaluation set embeddings, $\mathcal{N}_{t}(\mu_{t}, \Sigma_{t})$ is the target set embeddings, $tr$ is the trace of a matrix.
\item \textit{Instruction accuracy(IA)}. As mentioned in the main paper, we calculate IA according to the tag difference between edited music and target music. When edited music has tag set as $T_{e}$ and target music has tag set as $T_{t}$:
\begin{equation}
IA = \frac{1}{max(\left|T_{e}\right|, \left|T_{t}\right|)} {\left|T_{e} \cap T_{t}\right|}
\end{equation}
where $\left|\cdot\right|$ donate the number of set.

\item \textit{Chord Recognition Accuracy (CRA)}. The chord accuracy is defined as:
\begin{equation}
CA = \frac{1}{N_{t} * N_{c}}\sum_{i=1}^{N_{t}}\sum_{j=1}^{N_{c}} I\left\{C_{i,j} = C_{i,j}^{'}\right\}
\end{equation}
where $N_{t}$ is the number of tracks, $N_{c}$ is the number of chords, $C_{i,j}$ and $C_{i,j}^{'}$ are the $j$-th chord in the $i$-th target track and edited track, respectively.

\item \textit{Overlapped Area of Pitch Class Histogram and Inter-Onset Interval}. We first obtain histograms of pitch (divided into 12 categories) and IOIs (divided into 32 categories) according to their respective classification numbers, and convert the histograms into probability density functions (PDFs) using Gaussian kernel density estimation. We then compute the feature overlapped area(OA) between the PDFs of edited music and target music, the OA definition like:($\mathcal{D}_{OA}$)
\begin{equation}
\mathcal{D}_{OA}^{f} = \frac{1}{N_{t} * N_{b}}\sum_{i=1}^{N_{t}}\sum_{j=1}^{N_{b}} OA(\mathcal{P}_{i,j}^{f}, \hat{\mathcal{P}}_{i,j}^{f})
\end{equation}
where $f\in\left\{PCH, IOI\right\}$, $N_{t}$ is the number of tracks, $N_{b}$ is the number of bars, OA refers to the overlapping area of two distributions, $\mathcal{P}_{i,j}^{f}$ and $\hat{\mathcal{P}}_{i,j}^{f}$ are the distribution features in $j$-th bar and $i$-th track of edited music and target music. 
\end{itemize}

\subsection{Subjective Evaluation}

\begin{figure}[h]
\centering
\includegraphics[width=1.0\columnwidth]{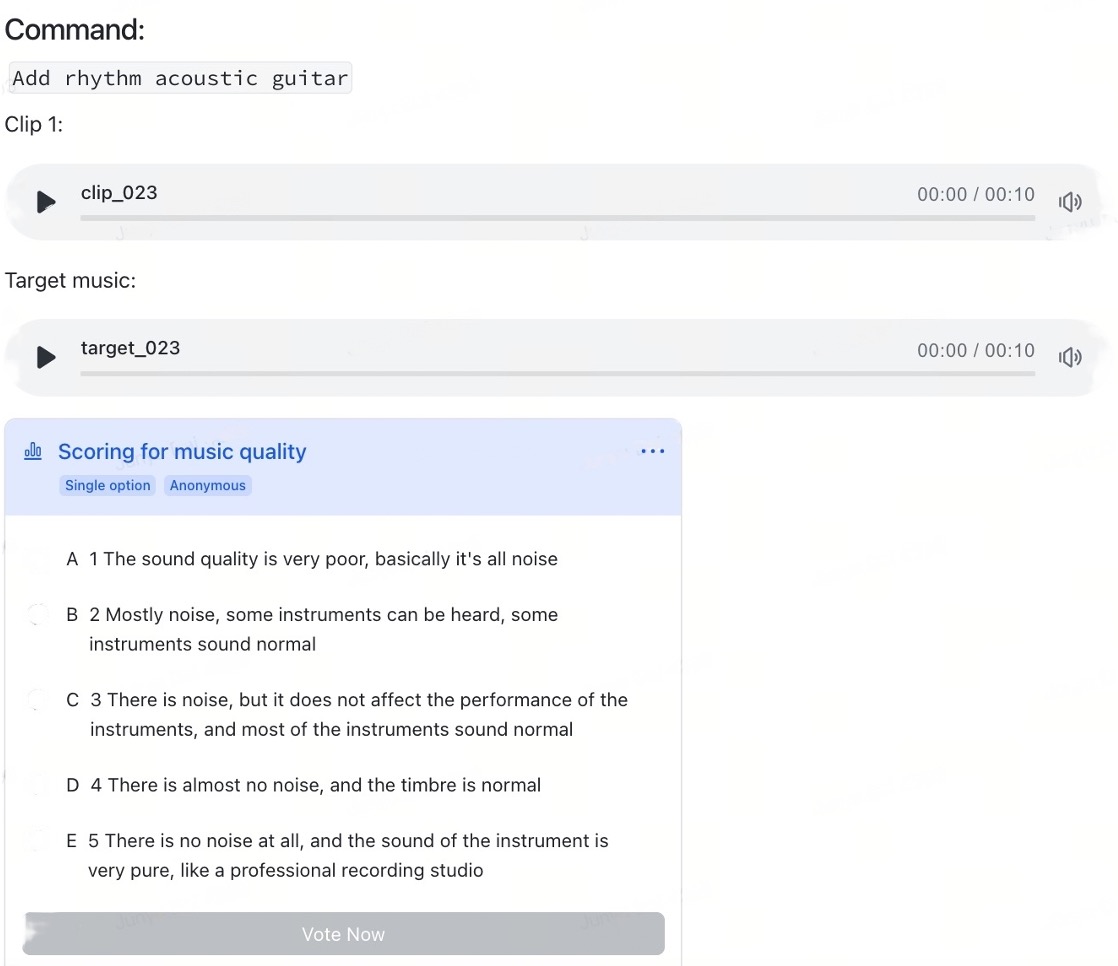} 
\caption{User interface of subjective evaluation. }
\label{fig:subject}
\end{figure}

Every candidate participant in the subjective evaluation has undergone standard alignment training to ensure that all participants score under the same standard. The training benchmark is as Table~\ref{tab:benchmark} shows. Figure~\ref{fig:subject} shows the user interface presented to candidates.

% \begin{table*}[ht]
% \centering
% %\resizebox{.95\columnwidth}{!}{
% \begin{tabular}{c|cc}
% \toprule
%     Quantitative Benchmark & \multicolumn{2}{c}{Music Quality} \\ \midrule
%     1  & \multicolumn{2}{p{1.6\columnwidth}}{The sound quality is very poor, basically it's all noise.} \\ \midrule
%     2 & \multicolumn{2}{p{1.6\columnwidth}}{Mostly noise, some instruments can be heard, some instruments sound normal.} \\ \midrule
%     3 & \multicolumn{2}{p{1.6\columnwidth}}{There is noise, but it does not affect the performance of the instruments, and most of the instruments sound normal.} \\ \midrule
%     4 & \multicolumn{2}{p{1.6\columnwidth}}{There is almost no noise, and the timbre is normal.} \\ \midrule
%     5 & \multicolumn{2}{p{1.6\columnwidth}}{There is no noise at all, and the sound of the instrument is very pure, like a professional recording studio.} \\ \bottomrule
% \end{tabular}
% \caption{Quantitative benchmark and Description of difference tasks.}
% \label{tab:command}
% \end{table*}

\subsection{Model Details}
\subsubsection{AutoEncoder}
Unlike image and text modalities, audio often has extremely low information density and ultra long sequences. For example, in 24khz sample rate, music with 10 seconds contains 240,000 samples, which is approximately 234 images in 32x32. To lower the computational demands of training diffusion models towards efficient music editing, we employ an autoencoder to learn a low dimension latent space which is perceptually equivalent to the waveform space, but offers significantly reduced computational complexity, like latent diffusion ~\cite{rombach2022high}.

Our autoencoder in this work is built based on a Variational AutoEncoder (VAE) model which consists of an encoder $\mathcal{E}$, a decoder $\mathcal{D}$ and a discriminator with 4 stacked ResNet-style~\cite{he2016deep} convolutional 1D blocks. Encoder $\mathcal{E}$ can compresses the waveform $x \in \mathbb{R}^{T\times 1}$ into a lower dimension latent space $z \in \mathbb{R}^{\frac{T}{r} \times C \times 1}$, where $f= r/C = 96/4 = 24$ can represent the compression ratio. Decoder $\mathcal{D}$ can reconstructs the waveform $\hat{x}$ from the latent space $z$. The discriminator is used in training process to distinguish between real and reconstructed waveform. Note that our autoencoder can reconstruct waveform directly, so there is no need for another vocoder like~\cite{wang2023audit,liu2023audioldm}. To train the VAE, we adopt a mean-square-error (MSE) based reconstruction loss $\mathcal{L}_{rec}$, an adversarial loss $\mathcal{L}_{adv}$, and a Kullback-Leibler loss $\mathcal{L}_{kl}$ to regularize the latent representation $z$. Then, the total training loss of VAE can be formulated as:
\begin{equation}
    \mathcal{L}_{VAE} = \mathcal{L}_{rec} + \mathcal{L}_{adv} + \mathcal{L}_{kl}
\end{equation}
The detail configuration of autoencoder are listed in Table~\ref{tab:autoencoder}. We pretrain autoencoder on the dataset we generated before for 2M steps with a learning rate of 2e-4 and batch size 16. And the optimizer we used is Adam.
\begin{table}[h]
    \centering
    \begin{tabular}{c|c}
    \toprule
    \multicolumn{2}{c}{AutoEncoder Configuration} \\ \midrule
    Number of Parameters     &  69.8M \\
    In/Out Channels & 1 \\
    % Latent Channels & 128 \\
    Number of Down/Up Blocks & 4 \\
    Cross Attention Dimension & 1024 \\
    Down sample out channels & [16, 32, 64, 128] \\
    Up sample out channels & [384, 192, 96, 48] \\
    Down sample Rate & [4, 4, 3, 2] \\
    Up sample Rate & [2, 3, 4, 4] \\
    \bottomrule
    \end{tabular}
    \caption{Configuration of AutoEncoder}
    \label{tab:autoencoder}
\end{table}

\subsubsection{Text Encoder}
To encode the text commands, we employ T5-large~\cite{raffel2020exploring}, as the text encoder $\mathcal{T}$ of InstructME, similar to~\cite{wang2023audit}. $\mathcal{T}$ is a pre-trained language model which is based on transformer and can convert text-based command into text embedding sequence for specifying the music editing task. It is important to note that the parameters of $\mathcal{T}$ is pre-trained and frozen during the training and sampling process. The detail is shown in Table~\ref{tab:t5}.

\begin{table}[h]
    \centering
    \begin{tabular}{c|c}
    \toprule
    \multicolumn{2}{c}{T5 Large Configuration} \\ \midrule
    Number of Parameters &  737.7M \\
    Output Channels  & 1024 \\
    \bottomrule
    \end{tabular}
    \caption{T5 Large}
    \label{tab:t5}
\end{table}

% \subsection{Chord Progression Extractor}

\subsubsection{U-Net}
We adopt the U-Net backbone of StableDiffusion~\cite{rombach2022high} as the basic architecture for InstructME. And the detail configuration is shown in Table~\ref{tab:unet}. 

All diffusion models in the experiments are trained using 16 NVIDIA TESLA V100 32GB GPUs with a batch size of 1 per GPU for 200k steps. We optimize them with the AdamW optimizer and initial learning rate 1e-4. 
\begin{table}[h]
    \centering
    \begin{tabular}{c|c}
    \toprule
    \multicolumn{2}{c}{Diffusion U-Net Configuration} \\ \midrule
    Number of Parameters&  898.9M \\
    In Channels & 1 \\
    Out Channels & 2 \\
    Learn Sigma Variance & True \\
    Number of Down/Up Blocks & 3 \\
    Number of ResBlocks & [4, 4, 4] \\
    Latent channels & [512, 1024, 1024] \\
    Down-Sample Rate & [2, 2, 2] \\
    Up-Sample Rate & [2, 2, 2] \\
    Number of Head Channels & 64 \\
    Cross Attention Dimension & 1024 \\
    % Down sample channels & [16, 32, 64, 128] \\
    % Down sample channels & [16, 32, 64, 128] \\
    % Down sample Rate & [4, 4, 3, 2] \\
    % Up sample Rate & [2, 3, 4, 4] \\
    \midrule
    Chunk Transformer Size & 200 \\
    \midrule
    Diffusion Steps & 1000 \\
    Noise Schedule & Linear \\
    Sampling Strategy & DDIM \\
    Sampling Steps & 200 \\
    \bottomrule
    \end{tabular}
    \caption{Configuration of Diffusion U-Net}
    \label{tab:unet}
\end{table}

\end{document}